\documentclass[12pt]{article}
\usepackage{axodraw}

  \newcommand{\ccaption}[2]{
    \begin{center}
    \parbox{0.85\textwidth}{
      \caption[#1]{\small{#2}}
      }
    \end{center}
    }

\newcommand{\DoT}{\left(\frac{\Delta}{2}\right)}
\newcommand{\lD}{\log \DoT}
\newcommand{\ltD}{\log^2 \DoT}

\newcommand{\ltpD}{\log^2\left(1+\frac{\Delta}{2}\right)}
\newcommand{\Li}{{\rm Li}_2}

\newcommand{\beq}{\begin{equation}}
\newcommand{\eeq}{\end{equation}}
\newcommand{\beqa}{\begin{eqnarray}}
\newcommand{\eeqa}{\end{eqnarray}}

\newcommand{\eqn}[1]{eq.~(\ref{#1})}
\newcommand{\eqns}[2]{eqs.~(\ref{#1}) and (\ref{#2})}
\newcommand{\Eqn}[1]{Eq.~(\ref{#1})}

\newcommand{\bq}{\bar{q}}
\newcommand{\cA}{{\cal A}}
\newcommand{\cO}{{\cal O}}
\newcommand{\cW}{{\cal W}}
\newcommand{\cM}{{\cal M}}
\newcommand{\dd}{{\rm d}}
\newcommand{\eps}{\epsilon}
\newcommand{\veps}{\varepsilon}

\newcommand{\la}{\langle}
\newcommand{\ra}{\rangle}

\newcommand{\tq}[1]{{\rm d} \widetilde{q}_{#1}}
\newcommand{\tk}[1]{{\rm d} \widetilde{k}_{#1}}
\newcommand{\tp}[1]{{\rm d} \widetilde{p}_{#1}}
\newcommand{\vp}[1]{\vec{p}_{#1}}
\newcommand{\vq}[1]{\vec{q}_{#1}}
\newcommand{\vk}[1]{\vec{k}_{#1}}
\newcommand{\slp}[1]{\!\not{\!p}_{#1}}
\newcommand{\slq}[1]{\!\not{\!q}_{#1}}

\newcommand{\slr}[1]{\!\not{\!r}_{#1}}
\newcommand{\sleps}{\!\!\not{\!\varepsilon}}

\newcommand{\ubar}{\bar{u}}
\newcommand{\vbar}{\bar{v}}

\newcommand{\op}[1]{\omega(\vp{#1})}
\newcommand{\oq}[1]{\omega(\vq{#1})}
\newcommand{\opq}{\omega(\vp1-\vq3)}

\begin{document}

\pagestyle{empty}
\begin{flushright}
  IPPP/03/67 \\
  DCPT/03/134 \\
\end{flushright}
\vspace*{1cm}
\begin{center}
  {\sc \large Infrared-Finite Amplitudes for Massless Gauge Theories} \\
   \vspace*{2cm}
{\bf Darren A. Forde} \ \
{\bf and} \ \
{\bf Adrian Signer}\\
\vspace{0.6cm}
{\it
Institute for Particle Physics Phenomenology \\
University of Durham \\
Durham, DH1 3LE, England \\ }
  \vspace*{2.8cm}
  {\bf Abstract} \\
\end{center}
\vspace*{5mm}
\noindent
We present a method to construct infrared-finite amplitudes for gauge
theories with massless fermions. Rather than computing $S$-matrix
elements between usual states of the Fock space we construct
order-by-order in perturbation theory dressed states that incorporate
all long-range interactions. The $S$-matrix elements between these
states are shown to be free from soft and collinear singularities. As
an explicit example we consider the process $e^+ e^-\to 2$ jets at
next-to-leading order in the strong coupling. We verify by explicit
calculation that the amplitudes are infrared finite and recover the
well-known result for the total cross section $e^+ e^-\to$ hadrons.

\newpage

\setcounter{page}{1}
\pagestyle{plain}


\section{Introduction}

Soft and collinear singularities which arise in gauge theories such as
QED and QCD are usually dealt with by summing up physically
indistinguishable cross sections \cite{Bloch:1937pw, KLN}. This
results in a cancellation of these singularities for so-called
infrared-safe observables and therefore reliable theoretical
predictions can be made for such observables. Even though this is a
perfectly valid approach to the problem it is useful to investigate
the origin of these singularities and to explore the possibility of
avoiding them altogether.

The origin of the problem lies in the long-range nature of the
interactions. As a consequence the usual in and out states do
not evolve in time asymptotically according to the free
Hamiltonian. It is this breakdown of the standard assumption that
results in the non-existence of the scattering operator. Thus, if we
want to avoid infrared singularities from the outset we have to
construct true asymptotic states and compute transition amplitudes
between them \cite{Chung, Zwanziger}.

This program has been carried out initially for QED with massive
fer\-mions~\cite{Kulish:ut} and then many steps have been made to
extend it to soft singularities in non-abelian
theories~\cite{Greco:1978te, Nelson, Ciafaloni}. It is possible to
construct asymptotic states (generalized coherent states) which
include multiple soft gluon emission to all orders in the
coupling~\cite{Catani:dp}. It can be shown that the $S$-matrix between
such states is free of soft singularities
\cite{Frenkel:gx,Giavarini:1987ts}. Apart from the more complicated
structure of the soft singularities due to the self-interaction of the
gauge bosons there is the additional complication of collinear
singularities in a non-abelian gauge theory. Due to the collinear
singularities the asymptotic Hamiltonian is more
complicated~\cite{Havemann:1985ra, DelDuca:jt, Contopanagos:1991yb,
Prokhorov:ri} and the prospect of being able to include these effects
to all orders in perturbation theory are not very promising. But the
idea of constructing an asymptotic Hamiltonian that takes into account
the asymptotic dynamics and using the corresponding evolution operator
to dress the usual states~\cite{Lavelle:1995ty, Bagan:1999jf} can
still be applied and is not tied to any particular theory. In
particular, four-point interactions that are present in non-abelian
gauge theories can be incorporated~\cite{Horan:1999ba}.

In the present work we investigate the practical feasibility of
constructing asymptotic states whose $S$-matrix elements are free from
soft and collinear singularities. We are not so much interested in
general considerations but rather try to establish a method to define
and explicitly compute infrared finite amplitudes order-by-order in
perturbation theory. Apart from the conceptional advantage of avoiding
divergent amplitudes such a method would have a variety of practical
advantages.  Obviously, the finiteness of the amplitudes would
facilitate a completely numerical approach to the calculation of
amplitudes. This also applies to the combination of fixed order
results with parton shower Monte Carlo programs.

After some general remarks about infrared singularities in
Section~\ref{infsing} we discuss how to construct the dressed states
in Section~\ref{asstates}. This discussion will be completely
general. In order to test the practicability of our method we consider
a simple process in Section~\ref{example}. We compute the infrared
finite amplitudes for $e^+e^-\to2$ jets at next-to-leading order in
the strong coupling and verify that upon integration over the phase
space they reproduce the well-known total cross section for
$e^+e^-\to$ hadrons. Technical details of this section are relegated
to an Appendix. Finally, in Section~\ref{summary}, we present a
summary and suggestions on how to improve the method to make
applications to more complicated processes more feasible.

\section{Infrared singularities \label{infsing}}

We are concerned with gauge theories with massless fermions. Such
theories are plagued by infrared singularities. Infrared singularities
are related to either arbitrarily soft gauge bosons or arbitrarily
collinear gauge bosons and/or fermions.\footnote{We use the term
``infrared singularities'' for both, soft and collinear singularities}
More precisely, if we define our external states in the usual way by
acting with creation operators on the vacuum, then higher-order
$S$-matrix elements between such external states contain infrared
divergences.

Many attempts have been made to define amplitudes that are well
defined, i.e.  do not contain such singularities. However, most
attempts were restricted to soft singularities. In particular it has
been shown that in an abelian gauge theory with massive fermions it is
possible to define external states whose $S$-matrix elements are free
from infrared singularities to all orders in perturbation
theory~\cite{Kulish:ut}.

An abelian gauge theory with massive fermions does not contain
collinear singularities. This simplifies the situation
considerably. As soon as we consider the non-abelian case, however, we
cannot avoid the appearance of collinear divergences. The reason is
that in the non-abelian case a massless gauge boson can split into two
arbitrarily collinear gauge bosons. Such a splitting results in a
collinear singularity. Thus, giving the fermions a mass does not
protect us from collinear singularities.

The aim of this work is to investigate the possibility of defining
infrared-finite amplitudes for a massless, non-abelian gauge
theory. The application we have in mind is, of course, QCD. In most
applications of perturbative QCD the quarks (at least the light
flavors) are treated as massless. Since we will have to deal with
collinear singularities anyway in a non-abelian theory, we might as
well take the common approach and treat the quarks as massless.

\subsection{The conventional approach \label{convap}}

Before we tackle the problem of defining infrared-finite amplitudes in
a situation where collinear singularities have to be taken into
account as well, it might be useful to remind ourselves how infrared
singularities are dealt with in the conventional approach.

In the conventional approach, which in the following we shall call the
cross-section approach, we evaluate amplitudes between conventional
external states that are obtained simply by acting with creation
operators on the vacuum. Since such amplitudes contain infrared
singularities we regularize them. In virtually all applications
dimensional regularization is used, whereby the amplitudes are
evaluated in $D\equiv 4-2\eps$ dimensions. The infrared singularities
then reveal themselves as poles in $1/\eps$.

In order to relate these amplitudes to measurable quantities, we will
have to integrate the (squared) amplitudes over the phase space,
weighted by a measurement function that defines the quantity we are
interested in. The key point is that we can only obtain theoretical
predictions for quantities that are infrared safe. An infrared safe
quantity is one that does not depend on whether or not a parton emits
an arbitrarily soft gluon. Also it must not depend on whether or not a
parton splits into two collinear partons. For such quantities the
infrared singularities present in the amplitudes are cancelled by
infrared singularities that appear due to the phase space integration
and after the cancellation the regulator can be removed. Even though
there are various general methods available to perform such
calculations at next-to-leading order~\cite{nlops} the appearance of
infrared singularities make the phase-space integration a non-trivial
problem.

\subsection{Asymptotic interactions}

One of the basic assumptions in the derivation of Greens functions or
transition amplitudes is that in the limit $t\to\pm\infty$ the
external states are free. Thus, it is assumed that all interactions
vanish rapidly enough so that they can be neglected in the distant
past and future. More precisely, we assume that for any exact state
vector $|\Psi(t)\rangle$ (in the Schr\"odinger picture) of the full
Hamiltonian $H$ we can find a corresponding state vector
$|\Phi(t)\rangle$ of the free Hamiltonian $H_0$ such that
 \beq
  |\Psi(t)\rangle = e^{-it H} |\Psi(0)\rangle\ \to \
  |\Phi(t)\rangle = e^{-it H_0} |\Phi(0)\rangle
 \label{basass}
 \eeq
for $t\to\pm\infty$. We then specify the in- and out-states $|\Phi_i
\rangle$ and $\langle \Phi_f|$ by a complete set of quantum numbers of
$H_0$ and compute the transition amplitudes
 \beq
 \langle \Phi_f|\Omega_-^\dagger \Omega_+|\Phi_i \rangle
 \equiv  \langle \Phi_f| S |\Phi_i \rangle
 \label{taSchr}
 \eeq
where we expressed the scattering operator $S$ in terms of the usual
M{\o}ller operators
 \beq
 \Omega_{\mp} \equiv \lim_{\tau\to\pm\infty}
 e^{i\tau H} e^{-i\tau H_0}
 \label{normMoll}
 \eeq
Since $S$ commutes with $H_0$ energy is conserved and the amplitude,
\eqn{taSchr}, is proportional to $\delta(E_i-E_f)$. This is a point we
will come back to below.

The important point is that in a gauge theory with conventional
external states the above mentioned assumption is simply
incorrect. The interactions due to massless gauge bosons are
long-range interactions and do {\it not} vanish rapidly enough for
$t\to\pm\infty$.  It is this violation of our basic assumption that
results in ill-defined, infrared-singular amplitudes. In other words,
given an exact state $|\Psi(t) \rangle$, in general there does not
exist a free state $|\Phi(t)\rangle $ such that \eqn{basass} is
satisfied.

If we want to tackle the problem at its root, we will have to compute
amplitudes between modified external states. The long-range
interactions that cause the problem will have to be included in the
external states themselves. This amounts to replacing our basic
assumption with the following: we can find an asymptotic Hamiltonian
$H_A$ such that for any exact state vector $|\Psi(t)\rangle$ we can
find an asymptotic state $|\Xi(t)\rangle$ such that
 \beq
 |\Psi(t)\rangle = e^{-it\, H} |\Psi(0)\rangle\ \to \
 |\Xi(t)\rangle = e^{-it\, H_A} |\Xi(0)\rangle
 \label{modass}
 \eeq
for $t\to\pm\infty$. We then define modified M{\o}ller operators
 \beq
 \Omega_{A \mp} \equiv \lim_{\tau\to\pm\infty}
 e^{i\tau H} e^{-i\tau H_A}
 \label{modMoll}
 \eeq
and compute modified $S$-matrix elements
 \beq
 \langle \Xi_f|\Omega_{A-}^\dagger \Omega_{A+}|\Xi_i \rangle
 \equiv  \langle \Xi_f| S_A |\Xi_i \rangle
 \label{tamodSchr}
 \eeq

Once all true long-range interactions are included in the definition
of the asymptotic Hamiltonian we are guaranteed that $S$-matrix
elements between such asymptotic states are free from infrared
singularities \cite{Frenkel:gx,Giavarini:1987ts}. These $S$-matrix
elements in turn are related to measurable quantities. In the same way
as several (squared) amplitudes contribute to a physical cross section
in the conventional approach, there will be several (squared)
$S$-matrix elements between asymptotic states contributing to an
observable. The crucial difference is that contrary to the
cross-section method all these contributions are separately finite.

To find the asymptotic Hamiltonian we have to split the interaction
Hamiltonian into a ``hard'' and a ``soft'' piece
  \beq
  H_I = H_H(\Delta) + H_S(\Delta)
  \label{Hsplit} \eeq
and define $H_A(\Delta)\equiv H_0+H_S(\Delta)$. The separation of the
Hamiltonian into two pieces is by no means unique. The only
requirement is that $H_S(\Delta)$ includes all true long-range
interactions. Thus, the emission of a soft gauge boson and the
splitting of a parton into two collinear partons has to be
included. But there is a lot of freedom on how precisely we make the
split between soft and hard interactions. To indicate this
arbitrariness we include the parameter $\Delta$ in the
notation. Later, we will often use the abbreviated notation
$H_\Delta\equiv H_S(\Delta)$. Let us stress that even though we will
call $H_\Delta\equiv H_S(\Delta)$ the soft Hamiltonian, it includes
all long-range interactions that potentially give rise to infrared
singularities. In particular, the soft Hamiltonian also includes the
splitting of a parton into two collinear partons.

In what follows we will define $H_A$ and the asymptotic states more
carefully. We then show how infrared-finite $S$-matrix elements are
related to conventional amplitudes. Finally, we consider a simple
explicit example, $e^+e^-\to$ 2 jets at next-to-leading order.

\section{Infrared-finite amplitudes \label{asstates}}

In order to obtain infrared-finite amplitudes we have to find true
asymptotic states and evaluate the (modified) $S$-matrix elements
between such states, as given in \eqn{tamodSchr}. Measurable cross
sections then are constructed out of these infrared-finite matrix
elements.

Since $S_A$ commutes with $H_A$ we conclude from \eqn{tamodSchr} that
energy is also conserved for the modified transition amplitudes.
However, given an asymptotic Hamiltonian it is generally not possible
to find the corresponding eigenstates $|\Xi_i\rangle$. First of all,
being an eigenstate of the asymptotic Hamiltonian which includes all
soft interactions, the true states $|\Xi_i\rangle$ would correspond to
quasi hadrons.  In our strict perturbative approach we will never be
able to describe bound states and, at each order in perturbation
theory, the states $|\Xi_i\rangle$ will correspond to some sort of
``jet-like'' states. In particular, these states are colored and will
have to be related to hadronic states using a hadronization
model. This is as in the cross-section approach and is an issue that
we do not address in this paper. Still, we have to relate the
asymptotic states $|\Xi_i\rangle$ to conventional free states
$|\Phi_i\rangle$ order by order in perturbation theory. We are then
led to compute matrix elements \beq \cM_{fi} \equiv \langle \Phi_f|
S_A |\Phi_i \rangle
  \label{mat2comp}
  \eeq 
order by order in perturbation theory and relate them to physical
(infrared safe) cross sections. It has been argued previously that
matrix elements as given in \eqn{mat2comp} are also free of soft
singularities~\cite{Frenkel:gx,Giavarini:1987ts}. In the present
article we extend this to include collinear singularities (see
also~\cite{Contopanagos:1991yb}). For more details on this issue we
refer to Section~\ref{sec:finS}.

\subsection{Notation and conventions}

Before we proceed let us fix our notation and conventions. Whereas
part of the discussion so far was done in the Schr\"odinger picture we
will now turn to the interaction picture. Thus all operators and
states are now to be understood to be given in the interaction
picture.

To start with we construct the usual states of the Fock
space
  \beq
  |q_i(p_i)\ldots \bq_j(p_j)\ldots g_k(p_k)\ldots \ra \equiv
  \prod_i b^\dagger(p_i) \, \prod_j d^\dagger(p_j)\,
  \prod_k a^\dagger(p_k)\, |0\ra
  \label{convstate} \eeq
where $b^\dagger,\, d^\dagger$ and $a^\dagger$ denote the creation
operators for fermions, antifermions and gauge bosons respectively and
we suppressed the helicity labels. We will generically denote such
states by $|i\ra$ and $\la f|$. Of course, we have to keep in mind
that the states as given in \eqn{convstate} are not normalizable and
we tacitly assume they have been smeared with test functions. Thus, we
are really concerned with wave packets. However, we assume they are
sharply peaked around a certain value of the momentum such as to
represent a particle beam with (nearly) uniform, sharp momentum.

The creation and annihilation operators satisfy the usual
(anti)com\-mu\-ta\-tion relations
\beqa
[a(\lambda_1, \vk1), a^\dagger(\lambda_2, \vk2)] &=&
-(2\pi)^3 2 \omega(\vk1) g_{\lambda_1 \lambda_2}\, \delta(\vk1-\vk2)
\label{acomm} \\
\{b(r_1, \vk1),b^\dagger(r_2,\vk2)\} &=&
(2\pi)^3 2 \omega(\vk1)\delta_{r_1 r_2} \,
\delta(\vk1-\vk2)
\label{bcomm} \\
\{d(r_1, \vk1), d^\dagger(r_2,\vk2)\} &=&
(2\pi)^3 2 \omega(\vk1)\delta_{r_1 r_2} \,
\delta(\vk1-\vk2)
\label{dcomm}
\eeqa
with $\omega(\vk{i}) \equiv |\vk{i}|$.  Note that the ordering used in
\eqn{convstate} implies a certain phase convention. Of course, all
amplitudes are only defined up to such a convention.

The field operators are given by
  \beqa
  \Psi_\alpha &=& \int \tk \,
  \left(u_\alpha(r,\vk{})\, b(r,\vk{})\, e^{-i k x} +
        v_\alpha(r,\vk{})\, d^\dagger(r,\vk{})\, e^{+i k x} \right)
  \label{psiop} \\
  \overline{\Psi}_\alpha &=& \int \tk \,
  \left(\ubar_\alpha(r,\vk{})\, b^\dagger(r,\vk{})\, e^{+i k x} +
        \vbar_\alpha(r,\vk{})\, d(r,\vk{})\, e^{-i k x} \right)
  \label{psibarop} \\
  A_\mu &=& \int \tk \,
  \left(\veps_\mu(\lambda,\vk{})\, a(\lambda,\vk{})\, e^{-i k x} +
        \veps_\mu^*(\lambda,\vk{})\, a^\dagger(\lambda,\vk{})\,
        e^{+ik x} \right)
  \label{Aop}
  \eeqa
where we defined
  \beq
  \tk{} \equiv \frac{d^{D-1} \vk{}}{(2\pi)^{D-1} 2\omega(\vk{})} \sum_{1,2}
  \label{sumS} \eeq
and the sum is over the two helicities of the fermions or gauge bosons
respectively.

Once the interaction Hamiltonian is given we can compute the evolution
operator and obtain in the interaction picture
  \beq
  U(t,t_0) \equiv
  T\, \exp\left(-i \int_{t_0}^{t} dt\, H_I(t) \right)
  \label{evU} \eeq
The M{\o}ller operators are given by $\Omega_\pm = U(0,\mp\infty)$ and,
thus, the scattering operator $S$ is related to the evolution operator
  \beq
  S = \Omega_-^\dagger \Omega_+ = U(+\infty,0) U(0, -\infty)
  \label{sop} \eeq
This allows us to find the $S$-matrix elements between some initial
and final state
  \beq
  \la f | S | i \ra =
  \la f | T\, \exp\left(-i \int_{-\infty}^{+\infty} dt\, H_I(t)
  \right) | i\ra
  \label{smat} \eeq
where $|i\ra$ and $\la f |$ are states as defined in
\eqn{convstate}. Inserting the explicit form of $H_I$ into \eqn{smat}
allows us to compute $S$-matrix elements. Of course, in practice such a
calculation is nothing but the computation of the corresponding
Feynman diagrams.

\subsection{Definition of infrared-finite amplitudes}

In analogy to \eqn{evU} we define a soft evolution operator
  \beq
  U_\Delta(t,t_0) \equiv
  T\, \exp\left(-i \int_{t_0}^{t} dt\, H_\Delta(t) \right)
  \label{softevU} \eeq
where we only include the soft Hamiltonian $H_\Delta(t)\equiv
H_S(\Delta,t)$. Acting on a certain state, the soft evolution operator
modifies this state by allowing for soft and collinear emissions.
Then, the usual Feynman-Dyson scattering matrix $S$ can be
decomposed as
  \beq
  S = U(+\infty,0) U(0, -\infty) \equiv
  \Omega_{\Delta -}^\dagger S_A(\Delta) \Omega_{\Delta +}
  \label{Ssplit} \eeq
where we have introduced the soft M{\o}ller operators
$\Omega_{\Delta\pm} \equiv U_\Delta(0,\mp\infty)$. More
explicitly, we have
  \beqa
  \Omega_{\Delta-}^\dagger &\equiv&
  T\, \exp\left(-i \int_0^{\infty} dt H_\Delta(t) \right)
  \label{Moll} \\
  &=& 1 -i  \int_0^{\infty} dt\, H_\Delta(t) +
  (-i)^2 \int_0^{\infty} dt\, \int_0^t dt'\, H_\Delta(t) H_\Delta(t')
  + \ldots \nonumber \\
  &=&\vspace*{-0.2cm} 1 -i \int_0^{\infty} dt\, H_\Delta(t) +
  \frac{(-i)^2}{2!} \int_0^{\infty} dt\, \int_0^{\infty} dt'\, T\{
  H_\Delta(t) H_\Delta(t') \}  + \ldots
  \nonumber \eeqa
\Eqn{Ssplit} defines a modified scattering operator $S_A(\Delta)$. This
operator has the crucial property that it includes at least one hard
interaction and, therefore, matrix elements $\la f| S_A(\Delta) |
i\ra$ of this operator with ordinary external initial and final states
as defined in \eqn{convstate} have no infrared singularities. If we
define dressed initial and final states, $|\{i\}\rangle$ and
$\langle\{f\}|$ according to
  \beqa
  |\{i\}\rangle &\equiv& \Omega_{\Delta+}^\dagger | i\rangle
  \label{iasdef} \\
  \langle\{f\}| &\equiv& \langle f|\Omega_{\Delta-}
  \label{fasdef}
  \eeqa
then
  \beq
  \langle\{f\}| S |\{i\}\rangle = \langle f|S_A(\Delta) | i\rangle
  \label{relSS} \eeq
Thus, the $S$-matrix elements of dressed states are free of
infrared singularities. We should stress that dressed states are not
asymptotic states, i.e. they are not eigenstates of the asymptotic
Hamiltonian.

Let us look at a dressed final state somewhat more carefully. We
obtain a dressed final state by acting with $\Omega_{\Delta-}$ on
a final state as defined in \eqn{convstate}. We denote this dressed
state by adding curly brackets.
  \beq
  _{f}\la \{ q(p_i)\ldots \bq(p_j)\ldots g(p_k)\ldots\} |
  = \la q(p_i)\ldots \bq(p_j)\ldots g(p_k)\ldots | \Omega_{\Delta-}
  \label{drstate} \eeq
Once the asymptotic Hamiltonian is fixed \eqn{drstate} is a unique
relation, order by order in perturbation theory, between an ordinary
final state $\langle f|$ and the corresponding dressed final state
$\langle\{f\}|$. A similar relation holds for dressed initial
states. 
  \beq
  |\{ q(p_i)\ldots \bq(p_j)\ldots g(p_k)\ldots\} \ra _{i}
  =\Omega^\dagger_{\Delta+} 
  |q(p_i)\ldots \bq(p_j)\ldots g(p_k)\ldots \ra 
  \label{drinstate} \eeq
In what follows we will suppress the labels $f$ and $i$ but keep in
mind that the states $|\{ q(p_i)\ldots \bq(p_j)\ldots g(p_k)\ldots\}
\ra$ and $\la \{ q(p_i)\ldots \bq(p_j)\ldots g(p_k)\ldots\} |$ are not
conjugates of each other. Also, we would like to stress that all these
states are states in the usual Fock space. Of course, this implicitly
assumes that we use some kind of regularization for the infrared
singularities in intermediate steps.

The soft M{\o}ller operators dress the usual non-interacting external
states with a cloud of soft and collinear partons. Since the infrared
behavior of $H_\Delta$ and the full interaction Hamiltonian are the
same by construction, this dressing generates infrared singularities
that cancel those generated by the full scattering operator.

There are two main differences between the soft(/collinear) M{\o}ller
operator, \eqn{Moll}, and the usual scattering operator,
\eqn{smat}. Firstly, $\Omega_{\Delta\pm}$ involve only the soft part
$H_\Delta$ of the interaction Hamiltonian. Secondly, the time
integration in the soft M{\o}ller operator runs only from $0$ to
$\infty$ rather than from $-\infty$ to $\infty$.

The fact that the time integration is restricted to $t>0$ is related
to the loss of Lorentz invariance in the amplitudes $\cM_{fi}$,
\eqn{mat2comp}. This is to be expected since $S_A$ does not commute
with $H_0$ and, therefore, $\cM_{fi}$ is generally not proportional to
an energy conserving $\delta(E_i-E_f)$. Instead, individual parts of
the amplitude will have $\delta$-functions with different energy
arguments (see \eqn{eq:SA}). The difference between these arguments
determines the amount by which energy conservation can be violated in
$\cM_{fi}$ and is related to the parameter $\Delta$. In the limit
$\Delta\to 0$ the amount by which energy can be violated tends to
0. Thus, the parameter $\Delta$ determines how much the initial wave
packets are distorted through the evolution with the soft M{\o}ller
operators. We will come back to these issues in Section~\ref{ampxs}.

\subsection{Factorization of modified $S$-matrix elements}

We now turn to the question on how to compute the infrared-finite
amplitudes defined in \eqn{relSS} and how they are related to
ordinary amplitudes.

A possible approach is to start from the right hand side of
\eqn{relSS}. This would involve using the explicit form of $S_A$,
given below in \eqn{eq:SA} to compute the amplitudes. As we argue in
Section~\ref{sec:finS} the structure of $S_A$ is such that no infrared
singularities occur. This opens up the possibility of evaluating the
amplitudes numerically. We have to keep in mind, however, that there
are still ultraviolet singularities which will have to be removed by
renormalization. In order to take an entirely numerical approach the
renormalization procedure would have to be done at the integrand
level~\cite{Nagy:2003qn}.

We will take a somewhat different approach in that we start from the
left hand side of \eqn{relSS}. We relate the infrared finite amplitude
to ordinary amplitudes by inserting a complete set of states twice
  \beq
  \la\{f\}| S |\{i\}\ra =
  \la f | \Omega_{\Delta-}\, S\, \Omega_{\Delta+}^\dagger |i \ra =
  \la f | \Omega_{\Delta-} | f' \ra \otimes
  \la f'| S | i'\ra \otimes
  \la i'| \Omega_{\Delta+}^\dagger |i \ra
  \label{facS} \eeq
Note that in the final expression all states are ordinary Fock space
states as defined in \eqn{convstate}. In this way, the infrared finite
amplitude is split into three pieces. First, there is an ordinary
$S$-matrix element, $\langle f'| S | i'\rangle$. The other two factors
are dressing factors for the initial and final state. All these pieces
are infrared divergent and only the complete amplitude is infrared
finite, order by order in perturbation theory. The ultraviolet
singularities appear only in $\langle f'| S | i'\rangle$ and are dealt
with as usual by renormalization. The symbol $\otimes$ denotes
integration over all momenta and summation over all helicities of the
state under consideration. Thus, for say $\langle f'| =
\langle q(p_1,r_1)\bq(p_2,r_2)g(p_3,r_3)| \equiv \langle
q_{p1}\bq_{p2}g_{p3}|$ we have
  \beq
  |f'\rangle \otimes  \langle f'| =
  \sum_{r_1\, r_2\, r_3} \int \tp1 \tp2 \tp3\
  |q_{p1} \bq_{p2} g_{p3}\rangle\,
  \langle q_{p1} \bq_{p2} g_{p3}|
 \label{otimesdef} \eeq
We should stress that \eqn{facS} implies that the dressing is not done
for each external parton separately. The dressing factors $ \langle f
| \Omega_{\Delta-} | f' \rangle$ do contain terms that factorize into
separate contributions for each parton, but they also contain color
correlated contributions.

\subsection{Dressing factors}

As we have seen in \eqn{facS} infrared-finite amplitudes are composed of three
factors. First, there is an ordinary amplitude, $\la f'| S | i'\ra$, computed
in the usual way using ordinary Feynman rules. Then there are the two dressing
factors, one for the initial and one for the final state. The calculation of
these dressing factors is somewhat different from the calculation of ordinary
amplitudes and it is useful to look at this in some more detail.

For concreteness we consider the calculation of a final state dressing factor.
The starting point is \eqn{Moll}. Let us stress again that since the time
integration in \eqn{facS} is from 0 to $\infty$ we break Lorentz invariance
right from the beginning. Of course, in the final result for a physical
quantity Lorentz invariance will be restored. In fact, the calculation has many
features of (old-fashioned) time-ordered perturbation theory. Most notably, all
particles will be on-shell. Three-momentum will be conserved in all vertices,
but energy will not be conserved.

A typical term of the (asymptotic) Hamiltonian that gives rise to an
$n$-point interaction has the form
  \beq
  \int \dd \vec{x}\, \int \prod_{i=1}^n \tk{i} \  V(\vk{i}) \Theta(\Delta)
   \, e^{ i  \vec{x}\cdot \sum \sigma_i \vk{i}} e^{- i t \sum \sigma_i
     \omega(\vk{i})}
  \label{eq:asH}
  \eeq
where $\omega(\vk{j}) \equiv |\vk{j}|$ denotes the energy of the particles and
the sign $\sigma_i$ is positive (negative) for incoming (outgoing) particles.
$V(\vk{i})$ is made up of creation and annihilation operators, eventually
accompanied by spinors and/or polarization vectors and a certain power of the
coupling constant. The range of integration over the momenta is restricted to
the singular regions. This is indicated in the notation by $\Theta(\Delta)$.
The precise form of this function is not important at the moment. After
performing the $\dd \vec{x}$ integration we obtain the momentum conserving
delta function $(2\pi)^{D-1} \delta^{(D-1)}(\sum \sigma_i \vk{i})$. However,
since the $t$ integration is restricted to $t\ge0$ we do not obtain an energy
conserving $\delta$ function. Rather we have to introduce the usual adiabatic
factor $0^+ > 0$ and use
 \beq
 \int_0^\infty \dd t\, e^{-i \omega t} \to
 \int_0^\infty \dd t\, e^{-i \omega t} e^{-t 0^+}
 = \frac{-i}{\omega - i 0^+}
 \label{tpos} \eeq
Of course, if the  $t$ integration was restricted to $t\le0$ we would
have
 \beq
 \int_{-\infty}^0 \dd t\, e^{-i \omega t} \to
 \int_{-\infty}^0 \dd t\, e^{-i \omega t} e^{+t 0^+}
 = \frac{i}{\omega + i 0^+}
 \label{tneg} \eeq
and the sum of \eqn{tpos} and \eqn{tneg} indeed results in $2\pi
\delta(\omega)$.

To summarize, for an $n$-point vertex in the calculation of a dressing factor
for a final state we have to use
 \beq
 \int \prod_{i=1}^n \tk{i}\, (2\pi)^{D-1}
 \delta^{(D-1)}(\sum \sigma_i \vk{i})\,
 \frac{\Theta(\Delta)}{\sum\sigma_i \omega(\vk{i}) - i 0^+} \,
 V(\vk{i})
 \label{FRdress} \eeq
Were it not for the $\Theta(\Delta)$ function and the restriction of
the $t$-integration to $t\ge0$ this would lead to the standard Feynman
rule.

\subsection{Finiteness of modified $S$-matrix elements \label{sec:finS}}

In this subsection we substantiate our claim that matrix elements as
defined in \eqn{mat2comp} or \eqn{relSS} are free from collinear and
soft singularities. We start from the definition
 \beq
 S_A(\Delta) = \Omega_{\Delta-} S\, \Omega_{\Delta+}^\dagger
 \label{SAexp}
 \eeq
and use the explicit form of the soft M{\o}ller operator and $S$ to
express $S_A(\Delta)$ in terms of $H_\Delta$ and $H_H$. Furthermore,
we observe that according  to \eqn{eq:asH} the time dependence of the
Hamiltonian $H(t_j)$ is given by
  \beq
  H_H(t_j) = h_j\, e^{-it_j\, \varpi_j}; \quad
  H_\Delta(t_j) = s_j\, e^{-it_j\, \varpi_j}
  \label{eq:Htdep}
  \eeq
where $ \varpi_j\equiv\sum \sigma_i \omega(\vk{i})$ is the sum of the
energies of the particles associated with the corresponding $n$-point
vertex and $h_j$ and $s_j$ are time independent. Performing the
algebra and the $t$-integrations we obtain up to third order
 \beqa
 \label{eq:SA}
 S_A &=&
 1 - 2 i \pi\, h_1\,  \delta(\varpi_1) \\
 &+& 2 i \pi\, h_1 h_2\,
 \frac{\delta(\varpi_1+\varpi_2)}{\varpi_1}
 + 2 i \pi\, \left[h_1, s_2\right]\,
 \frac{\delta(\varpi_1)-\delta(\varpi_1+\varpi_2)}{\varpi_2}
 \nonumber \\
 &+& 2 i \pi\, h_1 (h_2+s_2) h_3\,
 \frac{\delta(\varpi_1+\varpi_2+\varpi_3)}{\varpi_1 \varpi_3}
 \nonumber \\
 &-& 2 i \pi\, \left[\left[h_1,s_2\right]\!,s_3\right]\,
 \left(
 \frac{\delta(\varpi_1+\varpi_2)-
 \delta(\varpi_1+\varpi_2+\varpi_3)}{\varpi_1 \varpi_3} +
 \frac{\delta(\varpi_1)}{\varpi_2(\varpi_2+\varpi_3)}
 \right) \nonumber \\
 &-& 2 i \pi\, s_1 h_2 h_3\,  \frac{\delta(\varpi_2+\varpi_3)-
 \delta(\varpi_1+\varpi_2+\varpi_3)}{\varpi_1 \varpi_3}  \nonumber \\
 &-& 2 i \pi\, h_1 h_2 s_3\,  \frac{\delta(\varpi_1+\varpi_2)-
 \delta(\varpi_1+\varpi_2+\varpi_3)}{\varpi_1 \varpi_3}  \nonumber
 \eeqa
First of all we notice that all the purely soft terms $s_1 s_2 \ldots$
vanish. This holds to all orders and is crucial to ensure that $S_A$
is free from infrared singularities. Infrared singularities
potentially arise if $\varpi_i\to 0$. This corresponds to either a
soft or collinear emission at the corresponding vertex. Let us now go
through the terms in \eqn{eq:SA} and check that for none of them such
a singularity can occur. For this to be true we have to define $h_i$ such
that it vanishes for $\varpi_i \to 0$. This can be achieved by
choosing the $\Theta(\Delta)$ in \eqn{eq:asH} accordingly.

We start by looking at the second order terms, given in the second
line of \eqn{eq:SA}. The only potential singularity in the first term
is $\varpi_1\to 0$. This is harmless since $h_1=0$ in this
limit. In the second term we have the potential singularity
$\varpi_2\to 0$ which is not prevented by $s_2$. However, in this
limit the term is proportional to $\delta(\varpi_1)$ and the same
argument as for the first order term applies.

The arguments for the third order terms, given in the third to sixth
line of \eqn{eq:SA} are similar. The only dangerous limits in the
third line term for example are $\varpi_1\to 0$ and $\varpi_3\to
0$. Both of these are prevented by the presence of $h_1$ and
$h_3$. Considering the term in the fourth line, we first note that
$\delta(\varpi_1) h_1=0$. As a result there is no problem with the
limit $\varpi_2\to 0$ and $\varpi_2\to - \varpi_3$. Furthermore, the
singularity in the limit $\varpi_1\to 0$ is prevented by the presence
of $h_1$ and the limit $\varpi_3\to 0$ is made harmless by the
combination of $\delta$ functions. Similarly, the terms in the fifth
and sixth line are finite in the limit $\varpi_1\to 0$ and
$\varpi_3\to 0$. Thus we see that (up to this order) there are no
singularities in $S_A$ as long as $h_i$ is chosen to vanish for
$\varpi\to 0$.

We mention again that $S_A$ does not only contain terms proportional to
$\delta(\varpi_1+\varpi_2+\varpi_3)$ but also terms with ``incomplete''
$\delta$-functions. These are the energy violating terms mentioned above. We
also remark that the absence of terms containing $1/(\varpi_i+\varpi_j)$ in
$S_A$ (the corresponding term in the fourth line of \eqn{eq:SA} vanishes)
justifies our initial claim that all infrared singularities are related to
limits $\varpi_i\to 0$.

\subsection{Construction of infrared finite amplitudes}

The expression given in \eqn{facS} is a (double) sum over all possible
intermediate states $|f'\rangle \langle f'|$ and $|i'\rangle \langle i'|$.
However, if we compute the amplitude to a certain order in the coupling
constant, only a very limited number of intermediate states contribute. It is
for example clear that at order $\cO(g^0)$ the dressing factor $\langle f |
\Omega_{\Delta-} | f' \rangle$ is zero, unless $f=f'$. From this we
see that at leading order in perturbation theory the amplitude
$\langle\{f\}| S |\{i\}\rangle$ is the same as $\langle f| S |i\rangle$.

Including higher-order corrections this identity will, of course, not
hold any longer. At order $\cO(g^1)$ the states $f$ and $f'$ can be
different. To get a non-vanishing contribution they must be related
either by adding a (soft or collinear) gluon or by exchanging a
quark-antiquark pair by a gluon.

In order to illustrate this in more detail, let us consider a concrete
process. To simplify matters we consider a case with no partons in the
initial state. What we have in mind is for example the process $e^+
e^-\to \gamma \to$ jets. As long as we treat this process at leading
order in the electromagnetic coupling but at higher order in the
strong coupling, $g$, we encounter only final state
singularities. Thus, for the purpose of understanding how the dressing
removes the infrared singularities we can restrict ourselves to the
final state partons and treat the initial state simply as $|0\ra$.

Before writing down \eqn{facS} more explicitly for the process under
consideration, let us introduce a somewhat more compact notation. We
will denote the momenta and helicities of the partons in the
intermediate state $f'$ by $q_i$ and $s_i$ respectively and use the
notation $q_{qi}\equiv q(\vq{i},s_i)$ etc. The momenta and helicities
of the partons in the final state $f$ on the other hand will be
denoted by $p_i$ and $r_i$ and we use $q_{pi}\equiv q(\vp{i},r_i)$. The
order $\cO(g^n)$ terms of the dressing factors are then denoted by
  \beqa
  \lefteqn{g^n\, \cW^{(n)}(q_{p1}, \bq_{p2}, g_{p3} \ldots; q_{q1},
 \bq_{q2} \ldots) \equiv}   &&   \label{Wdef} \\
  && \la q(\vp1,r_1) \bq(\vp2,r_2) g(\vp3, r_3)\ldots|
  \Omega_{\Delta-} | q(\vq1,s_1) \bq(\vq2,s_2)\ldots \ra \Big|_{g^n}
  \nonumber \eeqa
Similarly, we denote the order $\cO(g^n)$ terms of the amplitude by
  \beqa
  \lefteqn{g^n\, \cA^{(n)}(q(\vq1,s_1), \bq(\vq2,s_2),
        g(\vq3,s_3)\ldots;\gamma) \equiv } && \label{Adef} \\
  && \la q(\vq1,s_1) \bq(\vq2,s_2) g(\vq3,s_3)\ldots|S|0\ra \Big|_{g^n}
  \equiv g^n\, \cA^{(n)}(q_{q1},\bq_{q2},g_{q3}\ldots;\gamma)
  \nonumber \eeqa
and we introduce a notation for the infrared finite amplitudes
  \beqa
  \lefteqn{g^n\, \cA^{(n)}(\{q_1(\vp1,r_1), \bq_2(\vp2,r_2),
        g_3(\vp3,r_3)\ldots\};\gamma) \equiv } && \label{AIRdef} \\
  && \la \{q(\vp1,r_1) \bq(\vp2,r_2) g(\vp3, r_3)\ldots\}|S|0\ra \Big|_{g^n}
  \equiv g^n\, \cA^{(n)}(\{q_{p1},\bq_{p2},g_{p3}\ldots\};\gamma)
  \nonumber \eeqa
We always make use of the convention that the helicity associated with
momentum  $\vp{i}$ is $r_i$ whereas the helicity associated with
momentum  $\vq{i}$ is $s_i$.

Let us now use \eqn{facS} to write down the infrared finite amplitude $\la
\{q(p_1,r_1)\bq(p_2,r_2)\}|S|0\ra$ order by order in perturbation theory. At
leading order we have
  \beqa
  \cA^{(0)}(\{q_{p1},\bq_{p2}\};\gamma) &\equiv&
  \la\{q(p_1,r_1)\bq(p_2,r_2)\}|S|0\ra \Big|_{g^0}  \label{exS0} \\
  &=& \cW^{(0)}(q_{p1},\bq_{p2};q_{q1},\bq_{q2}) \otimes
  \cA^{(0)}(q_{q1},\bq_{q2};\gamma) \nonumber \\
  &=& \cA^{(0)}(q_{p1},\bq_{p2};\gamma)
  \nonumber
  \eeqa
where in the last step we used
  \beqa
  \lefteqn{\cW^{(0)}(q_{p1},\bq_{p2}; q_{q1},\bq_{q2}) =} &&
  \label{exW0} \\
  &&
  (2\pi)^3 2 \omega(\vp1) \delta_{r_1 s_1} \delta(\vp1-\vq1)\
  (2\pi)^3 2 \omega(\vp2) \delta_{r_2 s_2} \delta(\vp2-\vq2)
  \nonumber \eeqa
\Eqn{exW0} is simply obtained by noting that $\Omega_{\Delta-} = 1$ at
$\cO(g^0)$, \eqn{Moll}, and using the (anti)commutation relations
eqs.(\ref{acomm}), (\ref{bcomm}) and (\ref{dcomm}).

At $\cO(g)$ the amplitude is zero because for every intermediate state
$f'$ either the dressing factor $\la f | \Omega_{\Delta-} | f' \ra$ or
the amplitude $\la f'| S | 0\ra$ vanishes.

At $\cO(g^2)$  the situation is more interesting. We have
  \beqa
  \cA^{(2)}(\{q_{p1},\bq_{p2}\};\gamma) &\equiv&
  \la\{q(p_1,r_1) \bq(p_2,r_2)\}|S|0\ra \Big|_{g^2}   \label{exS2}  \\
  &=& \cW^{(0)}(q_{p1},\bq_{p2};q_{q1},\bq_{q2}) \otimes
  \cA^{(2)}(q_{q1},\bq_{q2};\gamma) \nonumber \\
  &+& \cW^{(2)}(q_{p1},\bq_{p2};q_{q1},\bq_{q2}) \otimes
  \cA^{(0)}(q_{q1},\bq_{q2};\gamma) \nonumber \\
  &+& \cW^{(1)}(q_{p1},\bq_{p2};q_{q1},\bq_{q2},g_{q3}) \otimes
  \cA^{(1)}(q_{q1},\bq_{q2},g_{q3};\gamma) \nonumber
 \eeqa

 \begin{figure}[ht]
 \begin{picture}(300,100)(0,0)
 \Photon(300,50)(250,50){3}{5}
 \Line(240,50)(150,80)
 \LongArrow(216,58)(213,59)
 \Line(240,50)(150,20)
 \LongArrow(213,41)(216,42)
 \Line(150,75)(90,75)
 \Line(150,25)(90,25)
 \LongArrow(150,75)(110,75)
 \LongArrow(90,25)(110,25)
 \GOval(240,50)(15,15)(0){0.9}
 \GOval(150,50)(35,25)(0){0.9}
 \DashLine(200,80)(200,20){3}
 \Text(242,50)[c]{$\cA^{(i)}$}
 \Text(152,50)[c]{$\cW^{(j)}$}
 \Text(185,82)[c]{$q_1\ s_1$}
 \Text(185,18)[c]{$q_2\ s_2$}
 \Text(120,82)[c]{$p_1\ r_1$}
 \Text(120,18)[c]{$p_2\ r_2$}
 \end{picture}
 \ccaption{}{Cut diagrams for 2-particle intermediate state. The term
  $\cW^{(0)} \otimes \cA^{(2)}$ of \eqn{exS2T1} corresponds to $j=0,
  i=2$ and $\cW^{(2)} \otimes \cA^{(0)}$ corresponds to $j=2, i=0$.
  \label{fig:cut2p}}
 \end{figure}

The first term on the right hand side of \eqn{exS2} is nothing but the
usual one-loop amplitude multiplied by the $\cO(g^0)$ dressing factor
and, using \eqn{exW0}, can be written as
  \beq
  \cW^{(0)}(q_{p1},\bq_{p2};q_{q1},\bq_{q2}) \otimes
  \cA^{(2)}(q_{q1},\bq_{q2};\gamma)
  = \cA^{(2)}(q_{p1},\bq_{p2};\gamma)
  \label{exS2T1}
  \eeq
The second term is also a two-particle cut term, but this time it is the usual
tree-level amplitude multiplied by the next-to-leading order dressing factor.
These two terms are shown in Figure~\ref{fig:cut2p}.

 \begin{figure}[ht]
 \begin{picture}(300,100)(0,0)
 \Photon(300,50)(250,50){3}{5}
 \Gluon(240,50)(150,50){3}{9}
 \Line(240,50)(150,80)
 \LongArrow(216,58)(213,59)
 \Line(240,50)(150,20)
 \LongArrow(213,41)(216,42)
 \Line(150,75)(90,75)
 \Line(150,25)(90,25)
 \LongArrow(150,75)(110,75)
 \LongArrow(90,25)(110,25)
 \GOval(240,50)(15,15)(0){0.9}
 \GOval(150,50)(35,25)(0){0.9}
 \DashLine(200,80)(200,20){3}
 \Text(242,50)[c]{$\cA^{(1)}$}
 \Text(152,50)[c]{$\cW^{(1)}$}
 \Text(185,82)[c]{$q_1\ s_1$}
 \Text(185,18)[c]{$q_2\ s_2$}
 \Text(185,60)[c]{$q_3$}
 \Text(120,82)[c]{$p_1\ r_1$}
 \Text(120,18)[c]{$p_2\ r_2$}
 \end{picture}
 \ccaption{}{Cut diagram for 3-particle intermediate
 state. \label{fig:cut3p}}
 \end{figure}

The third term in \eqn{exS2} is of a somewhat different nature as it
is a three-particle cut diagram, as illustrated in
Figure~\ref{fig:cut3p}. The dressing factor
$\cW^{(1)}(q_{p1},\bq_{p2};q_{q1},\bq_{q2},g_{q3})$ is zero unless the
gluon $g_{q3}$ is either soft or collinear to the quark or
antiquark. Thus, the dressing factor projects out the infrared
singular piece of the bremsstrahlung amplitude. This is exactly the
piece that is needed to render the full amplitude
$\cA^{(2)}(\{q_{p1},\bq_{p2}\};\gamma)$ finite.

In the next section we will calculate this amplitude explicitly and check that
the infrared singularities present in the three terms of \eqn{exS2} cancel in
the sum.

The construction of the amplitude at higher orders in $g$ follows the same
pattern. For any odd power of $g$ the amplitude vanishes for the same reason as
it vanishes at $\cO(g)$. At $\cO(g^4)$ it is given by
  \beqa
  \lefteqn{\cA^{(4)}(\{q_{p1},\bq_{p2}\};\gamma) = } &&  \label{exS4} \\
  && \cW^{(0)}(q_{p1},\bq_{p2};q_{q1},\bq_{q2}) \otimes
  \cA^{(4)}(q_{q1},\bq_{q2};\gamma) \nonumber \\
  &+& \cW^{(2)}(q_{p1},\bq_{p2};q_{q1},\bq_{q2}) \otimes
  \cA^{(2)}(q_{q1},\bq_{q2};\gamma) \nonumber \\
  &+& \cW^{(4)}(q_{p1},\bq_{p2};q_{q1},\bq_{q2}) \otimes
  \cA^{(0)}(q_{q1},\bq_{q2};\gamma) \nonumber \\
  &+& \cW^{(1)}(q_{p1},\bq_{p2};q_{q1},\bq_{q2},g_{q3}) \otimes
  \cA^{(3)}(q_{q1},\bq_{q2},g_{q3};\gamma) \nonumber \\
  &+& \cW^{(3)}(q_{p1},\bq_{p2};q_{q1},\bq_{q2},g_{q3}) \otimes
  \cA^{(1)}(q_{q1},\bq_{q2},g_{q3};\gamma) \nonumber \\
  &+& \cW^{(2)}(q_{p1},\bq_{p2};q_{q1},\bq_{q2},g_{q3},g_{q4}) \otimes
  \cA^{(2)}(q_{q1},\bq_{q2},g_{q3},g_{q4};\gamma) \nonumber \\
  &+& \cW^{(2)}(q_{p1},\bq_{p2};q_{q1},\bq_{q2},q_{q3},\bq_{q4}) \otimes
  \cA^{(2)}(q_{q1},\bq_{q2},q_{q3},\bq_{q4};\gamma) \nonumber
 \eeqa
The separate terms in \eqn{exS4} are infrared divergent but in the sum
all these divergences cancel. This can be seen by looking at a
particular Feynman diagram, for example the one shown in
Figure~\ref{fig:cutXXp}, and realizing that \eqn{exS4} is nothing but the
sum over all possible cuts. Since the dressing factors are constructed
such that in the infrared limit they correspond to the usual
amplitudes it is clear that the infrared singularities in
$\cA^{(4)}(\{q_{p1},\bq_{p2}\};\gamma)$ have to cancel in the same way
as they cancel in ordinary cut diagrams. The first term of \eqn{exS4}
corresponds to the ordinary two-loop amplitude and is represented by
cut 1. The other two-particle cuts, the second and third term, are
represented by cut 2 and 3. There are two three-particle cut terms,
term 4 and 5. Finally, for the diagram under consideration, there is
one four-particle cut contribution, namely term 6. For a certain
Feynman diagram not all terms of \eqn{exS4} are present. In our case,
the last term of \eqn{exS4} which is another four-particle cut
contribution is missing.

 \begin{figure}[ht]
 \begin{center}
 \begin{picture}(200,130)(0,-10)
 \Photon(200,50)(150,50){3}{5}
 \Line(150,50)(0,100)
 \Line(150,50)(0,0)
 \GlueArc(95,83.4)(35,186,317){3}{10}
 \Gluon(30,90)(45,15){3}{9}
 \DashLine(38,100)(38,0){3}
 \DashLine(10,100)(10,0){3}
 \DashLine(90,100)(90,0){3}
 \DashLine(128,100)(128,0){3}
 \DashLine(80,100)(25,0){3}
 \DashLine(44,100)(64,0){3}
 \Text(10,-8)[c]{1}
 \Text(25,-8)[c]{6}
 \Text(38,-8)[c]{4}
 \Text(64,-8)[c]{2}
 \Text(90,-8)[c]{5}
 \Text(128,-8)[c]{3}
 \end{picture}
 \end{center}
 \ccaption{}{All possible cuts of a Feynman diagram representing
 the various terms of \eqn{exS4}.  \label{fig:cutXXp}}
 \end{figure}

We should stress that our approach to construct infrared finite
amplitudes is by no means restricted to amplitudes with final state
singularities only.  Initial state singularities are dealt with by
dressing the initial state, as can be seen in \eqn{facS}.

In fact, the dressing of the initial state would even be needed for
processes as discussed above, i.e. with say only a $\gamma$ in the
initial state. Above and in the rest of this paper we have excluded
any QED vertices from the soft Hamiltonian even though there is a
potential collinear singularity at this vertex. We do this because we
treat the incoming photon as off-shell and so it will generate no
infrared singularities.

If we were to include such vertices in the soft Hamiltonian then we
would generate many more diagrams with non-vanishing initial-state
dressing factors such as
$\cW(q(\vq1,s_1),\bq(\vq2,s_2),g(\vq3,s_3);\gamma)$. We would find
though, that all such extra contributes would cancel as all diagrams
with purely soft vertices cancel as described in
Section~\ref{sec:finS}.

\subsection{From amplitudes to cross sections \label{ampxs}}

Once the infrared finite amplitudes have been computed, they can be
used to compute cross sections for observables related to these
amplitudes. The procedure to obtain cross sections from amplitudes
depends to some extent on the external states we use and deserves some
further considerations.

In the cross-section approach we usually deal with amplitudes that are
proportional to a four-dimensional delta function. Upon taking the
absolute value squared, this leaves us with the problem of
interpreting the square of a delta function. Usually this is dealt
with in a rather non-rigorous manner by putting the system in a four
dimensional box of size $V\cdot T$. The square of the delta function
is then interpreted as $V\cdot T$ times a single delta function. The
factor $V\cdot T$ is cancelled by taking into account the
normalization of the states and the flux factor, which leaves us with
a cross section proportional to a single four-dimensional delta
function, expressing conservation of four momentum.

The appearance of the square of the delta function is of course
related to the fact that we usually work with non-normalizable states
with a sharp value of momentum and energy. In a more rigorous
treatment within the cross-section approach the in and out states
would have to be written as wave packets, sharply peaked around a
certain value of momentum and energy. It can then be shown that the
spreading of the wave packet during the scattering process can be
safely neglected~\cite{wavepack}. As mentioned in Section~\ref{convap}
the precise definition of the measurable quantity is given in terms of
a measurement function. This is a function of the partonic momenta. If
we are dealing with wave packets rather than sharp-momentum states,
the measurement function has to be defined in terms of these wave
packets. However, as long as we deal with wave packets whose spread is
well below any experimental resolution, we can simply use the normal
measurement function with the partonic momentum replaced by the
central value of the wave packet and we get the same result as in the
above mentioned, less rigorous approach~\cite{wavepack}. 

Let us now turn to the situation we encounter if we work with
infrared-finite amlitudes, defined in \eqn{mat2comp}. As mentioned
before, the amplitude is then not proportional to an energy conserving
delta function, even if we were to start with the usual
non-normalizable states. Following the proper treatment with wave
packets, we think of the states $|i\rangle$ and $\langle f|$ (or
$|\Phi_i\rangle$ and $\langle \Phi_f|$) as sharply peaked wave
packets. The states $|\{i\}\rangle$ and $\langle\{f\}|$ as defined in
\eqns{iasdef}{fasdef} are also wave packets. Through the action of the
M{\o}ller operators, their spread is larger than the spread of
$|i\rangle$ and $\langle f|$ and depends crucially on the parameter
$\Delta$. If we choose $\Delta$ small enough such that the spread of
the wave packets related to the states $|\{i\}\rangle$ and
$\langle\{f\}|$ is still smaller than any experimental resolution, we
can still compute any measurable cross section by using the standard
measurement function with the partonic momenta replaced by the central
value of the wave packet.

The important point is that we must be able to express any measurable
quantity in terms of the states $|\{i\}\rangle$ and
$\langle\{f\}|$. However, since the states $|\{i\}\rangle$ and
$\langle\{f\}|$ differ from $|i\rangle$ and $\langle f|$ only by soft
and collinear interactions, this is nothing but the requirement that
the quantity we are dealing with is infrared safe. Indeed, the
requirement of infrared safety states that the quantity must not
depend on wheter or not a parton emits another arbitrarily soft or
collinear parton. But in the limit $\Delta \to 0$ the soft M{\o}ller
operators do precisely this. Thus, choosing $\Delta$ small enough
ensures that the construction of the measurable quantity in terms of
the partonic momenta is not affected by the soft M{\o}ller
operators. 

This solves the problem on how to obtain differential cross sections,
once the infrared-finite amplitudes are known, in principle. In
practice, the explicit implementation of this programme is far from
trivial and requires further investigations. We mention for example
that choosing $\Delta$ very small might result in numerical problems,
similar to the so called binning problem in the standard approach. If,
on the other hand, we choose $\Delta$ too large (relative to the
experimental resoluion) the infrared-finite amplitudes are too
inclusive to allow the computation of any possible physical
quantity. It has been advocated before that the most convenient choice
of $H_A$ is the one that precisely corresponds to the experimental
resolution~\cite{Contopanagos:1991yb}. While this might be true in
principle, we think that this is not a practicable way to proceed,
since then the asymptotic Hamiltonian would depend on the details of
the experiment.

\section{An example $ e^+ e^-\to$ 2 jets at NLO \label{example}}

We consider the process $e^+ e^- \to \gamma(P) \to$ 2 jets. At leading
order there is only one partonic process that contributes, $e^+e^-\to
q\bq$. However, at next-to-leading order there is also the process
$e^+e^-\to q \bq g$. Since the initial state does not interact
strongly we can restrict our considerations to the process
$\gamma^*(P)\to$ 2 jets.

\subsection{The conventional approach}

In the conventional cross-section approach we compute the
amplitudes for the two partonic processes
  \beqa
  \lefteqn{\cA(q_{p1}, \bq_{p2}; \gamma(P)) =}&& \nonumber \\
    && \cA^{(0)}(q_{p1}, \bq_{p2}; \gamma(P))
     + g^2 \cA^{(2)}(q_{p1}, \bq_{p2}; \gamma(P)) + \cO(g^4)
  \label{sm0} \eeqa
and
  \beq
  \cA(q_{p1}, \bq_{p2}, g_{p3}; \gamma(P))
  = g \cA^{(1)}(q_{p1}, \bq_{p2}, g_{p3}; \gamma(P)) + \cO(g^3)
  \label{sm1}  \eeq
Upon squaring the amplitude and integration over the phase space we obtain
  \beq
  d \sigma = d \sigma_0 + g^2 d \sigma_{q\bq} + g^2 d \sigma_{q\bq g}
+ \cO(g^4)
  \label{smxs} \eeq
where
  \beqa
  d \sigma_0 &\sim& |\cA^{(0)}(q_{p1}, \bq_{p2}; \gamma(P))|^2
  \label{smsg0} \\
  d \sigma_{q\bq} &\sim& 2 {\rm Re}\left[ \cA^{(0)}(q_{p1}, \bq_{p2};
            \gamma(P))\cA^{(1)*}(q_{p1}, \bq_{p2}; \gamma(P)) \right]
            \label{smsgv} \\
  d \sigma_{q\bq g} &\sim& |\cA^{(1)}(q_{p1}, \bq_{p2}, g_{p3}; \gamma(P))|^2
  \label{smsgr}
  \eeqa
The virtual cross section, $d \sigma_{q \bq}$ and the real cross
section, $d \sigma_{q \bq g}$ both contain infrared singularities and
only when combined to form an infrared safe observable do these
divergences cancel. For the total cross section for example we obtain
  \beqa
  \sigma_0 &=& \frac{4\pi \alpha_{\rm em}^2 e_q^2 N_c}{3 s} \label{sig0} \\
  \sigma_{q\bq} &=& \sigma_0\, C_F \frac{\alpha_s}{\pi} c_\Gamma
        \left(\frac{2}{\epsilon^2} + \frac{3}{2\epsilon}
              +\frac{19}{4} - \frac{\pi^2}{2} \right) \label{sigv} \\
  \sigma_{q\bq g} &=&  \sigma_0\, C_F \frac{\alpha_s}{\pi} c_\Gamma
        \left(-\frac{2}{\epsilon^2} - \frac{3}{2\epsilon}
              - 4 + \frac{\pi^2}{2} \right)\label{sigr}
  \eeqa
where $c_\Gamma = 1 + \cO(\epsilon)$ and
$C_F=(N_c^2-1)/(2N_c)=4/3$. Thus, at next to leading order the total
cross section is given by
  \beq
  \sigma_1 = \sigma_{q\bq} + \sigma_{q\bq g} =
  \sigma_0 \left(1+\frac{\alpha_s}{4\pi}\, 3 C_F \right)
  \label{totXS} \eeq

\subsection{The infrared finite amplitudes}

In terms of infrared finite amplitudes, at next-to-leading order a cross
section is also made up of two contributions. The two amplitudes that
contribute are those with the final states $\la \{q_{p1} \bq_{p2}\}|$ and $\la
\{q_{p1} \bq_{p2} g_{p3}\}|$. However, the crucial point is that both these
amplitudes are infrared finite. Up to the order in $g$ required they are given
by
  \beqa
  \lefteqn{\cA(\{q_{p1}, \bq_{p2}\};\gamma) \equiv
  \la \{q_{p1} \bq_{p2}\}| S |0\ra =} \label{cs2part} \\
  && \int\, \tq1 \tq2\ \cW^{(0)}(q_{p1},\bq_{p2}; q_{q1},\bq_{q2}) \times
                   \cA^{(0)}(q_{q1},\bq_{q2}; \gamma(P)) \nonumber \\
  &+& \int\, \tq1 \tq2\ g^2\, \cW^{(0)}(q_{p1},\bq_{p2};
           q_{q1},\bq_{q2})
   \times
                   \cA^{(2)}(q_{q1},\bq_{q2}; \gamma(P)) \nonumber \\
  &+& \int\, \tq1 \tq2\ g^2\, \cW^{(2)}(q_{p1},\bq_{p2};
           q_{q1},\bq_{q2})
  \times
                   \cA^{(0)}(q_{q1},\bq_{q2}; \gamma(P)) \nonumber \\
  &+& \int\, \tq1 \tq2 \tq3\ g^2\,
  \cW^{(1)}(q_{p1},\bq_{p2}; q_{q1},\bq_{q2},g_{q3})
    \nonumber \\
  && \hspace*{5cm}
  \times\ \cA^{(1)}(q_{q1},\bq_{q2},g_{q3}; \gamma(P))  \ + \ \cO(g^4)
  \nonumber
  \eeqa
and
  \beqa
  \lefteqn{\cA(\{q_{p1}, \bq_{p2}, g_{p3}\};\gamma) \equiv
  \la \{q_{p1} \bq_{p2} g_{p3}\}| S |0\ra =} \label{cs3part} \\
  && \int\, \tq1 \tq2\ g\, \cW^{(1)}(q_{p1},\bq_{p2}, g_{p3}; q_{q1},\bq_{q2})
   \times   \cA^{(0)}(q_{q1},\bq_{q2}; \gamma(P)) \nonumber \\
  &+& \int\, \tq1 \tq2 \tq3 \ g\,
      \cW^{(0)}(q_{p1},\bq_{p2},g_{p3}; q_{q1},\bq_{q2},g_{q3}) \nonumber \\
  && \hspace*{5cm} \times \
      \cA^{(1)}(q_{q1},\bq_{q2},g_{q3}; \gamma(P)) \ + \ \cO(g^3) \nonumber
  \eeqa
where a sum over the spin/helicities of the intermediate
particles is understood to be included in $\int\tq{i}$, \eqn{sumS}.

\subsection{The asymptotic Hamiltonian}

Before we can proceed with the calculation of the infrared finite amplitudes we
have to define the asymptotic Hamiltonian $H_\Delta$. Once we have $H_\Delta$
we can obtain the M{\o}ller operator, \eqn{Moll}, and use it to construct the
dressed states, \eqn{drstate}, order by order in perturbation theory.

The only condition on $H_\Delta$ is that it includes all long-range
interactions from the original Hamiltonian. In order to separate these long
range soft and collinear emission terms from the hard emission terms we need to
introduce (at least) one parameter which we denote by $\Delta$. The dependence
of the asymptotic Hamiltonian on this parameter is indicated in the notation
$H_\Delta$. Once these terms are incorporated into the asymptotic Hamiltonian
we are free to include any other terms from the original Hamiltonian that we
wish, as these will only produce finite $\Delta$-dependent contributions to the
two final amplitudes. It is clear from \eqn{facS} that for the final result
this $\Delta$ dependence has to cancel.

In our case the only term of the interaction Hamiltonian we wish to include
in $H_\Delta$ is the quark gluon interaction vertex,
  \beq
  H_I \equiv g\, \int d\vec{x}\, :\overline{\Psi} T^a\, \gamma^\mu \Psi:
  A_\mu^a
  \label{Hint} \eeq
Using eqs.~(\ref{psiop},\ref{psibarop}) and (\ref{Aop}) we see that $H_I$
consists of eight terms
  \beqa
  H_I &=& g\,T^a\, \int \tk1 \tk2 \tk3\,
  \sum_{i=1}^8 V_i(\vk1,\vk2,\vk3)\label{V8def} \\
  && \hspace*{2cm}\times\,
  \exp\left(-i  t \sum_{j=1}^{3} \sigma_{i j}
  \omega(\vk{j})\right)
  \delta^{(D-1)}\!\left( \sum_{j=1}^{3} \sigma_{i j} \vk{j}\right),
  \nonumber \eeqa
where (suppressing the helicity and color labels)
  \beqa
  V_1 &=& b^\dagger(\vk1) b(\vk2) a(\vk3) \cdot
       \ubar(\vk1) \sleps(\vk3) u(\vk2),
       \nonumber \\
  V_2 &=& b^\dagger(\vk1) d^\dagger(\vk2) a(\vk3) \cdot
       \ubar(\vk1) \sleps(\vk3) v(\vk2),
       \nonumber \\
  V_3 &=& d(\vk1) b(\vk2) a(\vk3) \cdot
       \vbar(\vk1) \sleps(\vk3) u(\vk2),
       \nonumber \\
  V_4 &=& -d^\dagger(\vk1) d(\vk2) a(\vk3) \cdot
       \vbar(\vk2) \sleps(\vk3) v(\vk1),
       \nonumber \\
  V_5 &=& b^\dagger(\vk1) b(\vk2) a^\dagger(\vk3) \cdot
       \ubar(\vk1) \sleps^*(\vk3) u(\vk2),
       \nonumber \\
  V_6 &=& d(\vk1) b(\vk2) a^\dagger(\vk3) \cdot
       \vbar(\vk1) \sleps^*(\vk3) u(\vk2),
       \nonumber \\
  V_7 &=& b^\dagger(\vk1) d^\dagger(\vk2) a^\dagger(\vk3) \cdot
       \ubar(\vk1) \sleps^*(\vk3) v(\vk2),
       \nonumber \\
  V_8 &=& -d^\dagger(\vk1) d(\vk2) a^\dagger(\vk3) \cdot
       \vbar(\vk2) \sleps^*(\vk3) v(\vk1)
       \label{vsdef}
  \eeqa
The sign factors $\sigma_{i j}$ are $+1\ (-1)$ for incoming
(outgoing) particles.

As we only have to include terms in $H_{\Delta}$ that contribute in
the singular regions we are free to exclude the $V_i$ in \eqn{vsdef}
for which $\sum\sigma_i \omega(\vk{i})$ can never equal zero with all
$\omega(\vk{i})\ge0$ such that not all of the $\omega(\vk{i})=0$.  We
can see from \eqn{FRdress} that such terms will always be finite. From
the remaining terms we choose only those that give a singularity in
the physically relevant soft or collinear regions. This means that for
our example we can exclude $V_2$ and $V_6$, as these only go singular
when the two incoming or outgoing quarks from the vertex are
collinear. We emphasize that for more general processes these terms
have to be included in the asymptotic Hamiltonian.

We can confine the remaining terms even further as we are free to choose the
form of the finite part of $H_\Delta$. We restrict the integration of the
momenta $\vk1, \vk2$ and $\vk3$ to just the potentially singular regions. This
restriction is achieved here by including a theta function,
$\Theta(\Delta_i(\vk1,\vk2,\vk3))$ in each $V_i$ from \eqn{vsdef} which will
appear in $H_{\Delta}$. The form of $\Delta_i(\vk1,\vk2,\vk3)$ is completely
arbitrary as long as $\Theta \rightarrow 1$ in the soft and collinear limits.

The form of the $\Theta$ function that we will take for this example is,
  \beq
  \Theta(\Delta_i(\vk1,\vk2,\vk3)) \equiv
  \Theta(\Delta-|\sum_j \sigma_{ij} \omega(\vk{j})|).
  \label{thetadef} \eeq
This choice of $\Delta(\vk1,\vk2,\vk3)$ is particularly appropriate
because as we see in \eqn{FRdress}, $\sum_j \sigma_{ij}
\omega(\vk{j})$ is the exact form that the singular terms take. This
theta function therefore restricts the integral to just the regions
close to these singular limits.

By splitting up the covariant vertex into pieces and restricting the
integration to just the singular regions we are removing the manifest
Lorentz and gauge invariance from the amplitudes. Physical observables
will though be Lorentz and gauge invariant as we are effectively just
performing a unitary transformation (as we have regulated the
$\Omega_{\pm}$ operators) on a known Lorentz and gauge invariant
result.

To summarize, for our asymptotic Hamiltonian we take just the vertices $V_1$,
$V_4$, $V_5$ and $V_8$, giving,
  \beqa
  H_\Delta &=& g\, \int \tk1 \tk2 \tk3\,
  \sum_{i=1,4,5,8}\Bigg\{ V_i(\vk1,\vk2,\vk3)\exp\left(-i  t \sum_{j=1}^{3}
  \sigma_{ij} \omega(\vk{j})\right) \nonumber \\
  && \delta^{(D-1)}\!\left( \sum_{j=1}^{3} \sigma_{ij} \vk{j}\right)
  \Theta(\Delta-|\sum_j \sigma_{ij} \omega(\vk{j})|)\Bigg\}.
  \label{Hasymee} \eeqa

\subsection{Diagrammatic rules for the asymptotic regions
\label{sec:Feynman_Like_Rules}}

We can now take \eqn{Hasymee} and use it in \eqn{Moll} to form the
asymptotic operator. We could then go on to calculate the dressed
states of \eqn{Wdef} with this operator defined between suitable in
and out states by using the commutation relations between $a$, $b$,
$d$, $a^{\dag}$, $b^{\dag}$, $d^{\dag}$ and time-ordered perturbation
theory. However it can be shown that there are a set of diagrammatic
rules for the asymptotic region which behave in a similar way to
Feynman diagrams in normal perturbative field theory. Using these we
rules we can simplify the calculation.

These diagrammatic rules consist of vertex and propagator `like'
objects, but unlike normal Feynman diagrams we must take all time
orderings of the vertices into account. This is because we base the
evaluation of the amplitude in the asymptotic region on time ordered
perturbation theory. As mentioned before energy is not conserved at
each vertex and since the range of the time integration in the
M{\o}ller operators is from $0$ to $\infty$ there is no overall energy
conservation.

As there is a time ordering to the vertices we have both absorption
and emission rules. These are defined in
Figure~\ref{fig:Feynman_Vertex} with time flowing from right to left.

\begin{figure}[htb]
        \begin{center}
                \begin{minipage}{0.2\textwidth}
              \begin{picture}(60,60)(0,0)
                \Photon(25,12)(50,37){3}{5}
                \LongArrow(60,12)(45,12)
                \LongArrow(45,12)(12,12)
                \Line(12,12)(0,12)
                \LongArrow(30,30)(22,22)
                \Vertex(25,12){1.5}
                \Text(45,4)[c]{$p_3$}
                \Text(36,36)[c]{$p_2$}
                \Text(13,4)[c]{$p_1$}
                \Text(55,42)[c]{$\mu, a$}
              \end{picture}
                \end{minipage}%
                \begin{minipage}{0.6\textwidth}
          \beqa
                && \equiv\ (i g)T^a \gamma^{\mu}
                \frac{\delta^{(3)}(\vp{3}+\vp{2}-\vp{1})}
                  {\omega(\vp{3})+\omega(\vp{2})-\omega(\vp{1})-i0^+}
                  \nonumber \\
          && \hspace*{1cm} \times \
          \Theta(\Delta-|\omega(\vp{3})+\omega(\vp{2})-\omega(\vp{1})|)
          \nonumber \eeqa
                \end{minipage}
                \begin{minipage}{0.2\textwidth}
              \begin{picture}(60,60)(0,0)
                \Photon(35,12)(10,37){-3}{5}
                \LongArrow(60,12)(47,12)
                \LongArrow(47,12)(15,12)
                \Line(22,12)(0,12)
                \Vertex(35,12){1.5}
                \LongArrow(33,28)(25,36)
                \Text(48,4)[c]{$p_3$}
                \Text(39,24)[c]{$p_2$}
                \Text(15,4)[c]{$p_1$}
                \Text(5,42)[c]{$\mu, a$}
              \end{picture}
                \end{minipage}%
                \begin{minipage}{0.6\textwidth}
          \beqa
            &&  \equiv(-i g)T^a \gamma^{\mu}
            \frac{\delta^{(3)}(\vp{1}+\vp{2}-\vp{3})}
                  {\omega(\vp{1})+\omega(\vp{2})-\omega(\vp{3})-i0^+}
            \nonumber \\
             && \hspace*{1cm} \times \
            \Theta(\Delta-|\omega(\vp1{})+\omega(\vp{2})-\omega(\vp{3})|)
            \nonumber  \eeqa
                \end{minipage}
        \caption{The diagrammatic rules for vertices.}
        \label{fig:Feynman_Vertex}
        \end{center}
\end{figure}

We form propagator `like' objects from the spin sums of fermion
spinors and an associated energy denominator. Although they are not
real propagators in the normal field theory sense of inverted
off-shell two-point Green functions, they do represent the transition
from one vertex to another. The rules for these are shown in
Figure~\ref{fig:Feynman_Propagator}, where $\overline{p}^\mu\equiv
(p^0, -\vec{p})$.

\begin{figure}[!htb]
        \begin{center}
                \vspace*{-0.5cm}
                \begin{minipage}{0.2\textwidth}
              \begin{picture}(50,25)(0,0)
                \LongArrow(45,10)(25,10)
                \Line(25,10)(7,10)
                \Text(28,17)[c]{$p$}
              \end{picture}
                \end{minipage}%
                \begin{minipage}{0.5\textwidth}
          \beqa
                       \hspace*{-4.25cm}
      \equiv \frac{i \slp{}}{2 \omega(\vp{} )}. \nonumber
          \eeqa
                \end{minipage}
                \begin{minipage}{0.2\textwidth}
              \begin{picture}(50,25)(0,0)
                \Photon(7,10)(45,10){3}{5}
                \LongArrow(28,18)(15,18)
                \Text(34,19)[c]{$p$}
                \Text(0,15)[c]{$\mu$}
                \Text(0,5)[c]{$a$}
                \Text(55,15)[c]{$\nu$}
                \Text(55,5)[c]{$b$}
              \end{picture}
                \end{minipage}%
                \begin{minipage}{0.5\textwidth}
          \begin{flushleft}
            \beqa
                     \equiv \frac{\delta^{a b}}{2 \omega(\vp{} )}
                    \left( -g_{\mu \nu}+
                     \frac{p_{\mu}\overline{p}_{\nu}+
                     p_{\nu}\overline{p}_{\mu}}{
                     (p\overline{p})}\right).
              \nonumber
            \eeqa
          \end{flushleft}
                \end{minipage}
        \caption{The diagrammatic rules for propagators.}
        \label{fig:Feynman_Propagator}
        \end{center}
\end{figure}

As with ordinary field theory we must integrate over all internal momenta and
so for each propagator in the asymptotic region we must integrate over its
momentum $\int d^{D-1}p/(2\pi)^{D-1}$ in $D-1=3-2\epsilon$ dimensions. The
rules for external particles are exactly the same as for QED or QCD and so do
not need to be reproduced here. Finally we must include a factor of $1/n!$ with
each diagram, where $n$ represents the order in the coupling in the asymptotic
region.

As stated before the soft M{\o}ller operators are not necessarily
gauge invariant and neither Lorentz invariant. Infrared singularities
though will only occur in the region where $\varpi= \sum\sigma_i
\omega(\vk{i})=0$.  In this limit Lorentz invariance is restored and
so the structure of the singularities will also be Lorentz
invariant. Given that our amplitudes will not be gauge invariant, we
will perform all calculations including the second term of the gluon
propagator which ensures that we sum over physical polarizations only.

\subsection{The amplitude $\cA(\{q(p_1), \bq(p_2)\};\gamma)$ \label{sec:aqq}}

Let us start with the amplitude $\cA(\{q_{p1}, \bq_{p2}\};\gamma)$
given in \eqn{cs2part}. This amplitude consists of four terms and we
will look at each of them in turn. The first and second term will be
dealt with in Section~\ref{sec:born} and \ref{sec:virt}
respectively. For the third term of \eqn{cs2part} we need the dressing
factor $\cW^{(2)}(q_{p1},\bq_{p2}; q_{q1},\bq_{q2})$. There are
various combinations of interaction terms of the asymptotic
Hamiltonian that give rise to non-vanishing contributions to
$\cW^{(2)}$. These can be found using the diagrammatic rules of
Section~\ref{sec:Feynman_Like_Rules}. We find that there are four
contributing diagrams. These four diagrams can be split into two
classes, two self-interaction terms and two one-gluon exchange terms. We
consider the former in Section~\ref{sec:si} and the latter in
Section~\ref{sec:1_gluon_ext}. Finally, the last term of \eqn{cs2part}
will be computed in Section~\ref{sec:tpcd}.

\subsubsection{The Born term \label{sec:born}}

The first term is
  \beq
  \int\, \tq1 \tq2\ \cW^{(0)}(q_{p1},\bq_{p2}; q_{q1},\bq_{q2}) \times
                 \cA^{(0)}(q_{q1},\bq_{q2}; \gamma(P))
  \label{term1}
  \eeq
and is of order $g^0$. As discussed previously, \eqn{exS0}, this term
corresponds precisely to the tree-level amplitude.
 \beqa
 \cA^{(0)}(\{q_{p1},\bq_{p2}\}; \gamma(P)) &=&
 (-i e)\,\delta_{ij}\,\la p_1| \gamma^\alpha | p_2 \ra\, (2\pi)^D
 \delta^{(D)} (P-p_1-p_2) \nonumber \\
 &=&\cA^{(0)}(q_{p1},\bq_{p2}; \gamma(P)). \nonumber
 \label{cs2part1}
 \eeqa
We use a notation where $\la p_i|$ represents the spinor of a massless
outgoing fermion with momentum $p_i$ and similarly $| p_j \ra$
represents the spinor of a massless incoming fermion of momentum
$p_j$. Of course, these spinors depend on the helicity of the fermion,
but we suppress this dependence in the notation. The delta function as
usual ensures energy-momentum conservation for the process and the
$\delta_{ij}$ represents the color flow through the diagram.

\subsubsection{The virtual term \label{sec:virt}}

In the same way we see that the second term of \eqn{cs2part} corresponds to the
one-loop amplitude
  \beqa
 \lefteqn{\cA^{(2)}(q_{p1},\bq_{p2}; \gamma(P)) = } &&
 \label{cs2part2}  \\
 && C_F\left(\frac{\alpha_{s}}{2\pi}\right)
 \left(\frac{\mu^2}{s}\right)^{\epsilon}
 \left(-\frac{1}{\eps^2} - \frac{3}{2\eps} -
 4 + \frac{c_R}{2}+\frac{\pi^{2}}{12}\right) \,
 \cA^{(0)}(q_{p1},\bq_{p2}; \gamma(P)).
 \nonumber \eeqa
The infrared singularities appearing in \eqn{cs2part2} will be cancelled by
infrared singularities of the third and fourth term of \eqn{cs2part}. We should
mention that the finite term in \eqn{cs2part2} depends on the regularization
scheme used. The result in conventional dimensional regularization is obtained
by setting $c_R=0$ whereas in dimensional reduction we set $c_R=1$.

\subsubsection{The self-interaction terms \label{sec:si}}

The two self-interaction terms are obtained by taking the interacting
terms $V_1, V_5$ and $V_4, V_8$ of the asymptotic Hamiltonian as given
in \eqn{vsdef}. Since there is a symmetry between these two
contributions we only need to calculate one of the pair of
diagrams. The self interacting term resulting from the vertices $V_1,
V_5$ is shown in Figure~\ref{fig:2ptselfint} and is given by
  \beqa
  a_{15}^{\{2,0\}} &\equiv&
  \int \tq1 \tq2\ g^2\ \cW^{(2)}_{15}(q_{p1},\bq_{p2}; q_{q1},\bq_{q2})
  \times \cA^{(0)}(q_{q1},\bq_{q2}; \gamma(P))\label{W15a} \\
  &=&-\frac{(-i e)\,g^2}{2} \int d^{D-1}q_1 d^{D-1}q_2
  \frac{d^{D-1}q_3}{(2\pi)^{D-1}}
  \,\, T^a_{ik} T^b_{kj}\nonumber \\
  &&\frac{\delta^{ab}}{2\omega(\vq{3})}
  \left(-g^{\mu \nu}+\frac{q_3^{\mu}\,\overline{q}_3^{\nu}+q_3^{\nu}
  \,\overline{q}_3^{\mu}}{(q_3 \overline{q}_3)} \right)
  \la p_1| \gamma_{\mu} \slq{2} \gamma_{\nu} \slq{1}
  \gamma^{\alpha}|p_2\ra
  \nonumber \\
  &&\frac{\Theta(\Delta-|\omega(\vq{2})+\omega(\vq{3})-\omega(\vp{1})|)\,
  \Theta(\Delta-|\omega(\vq{2})+\omega(\vq{3})-\omega(\vq{1})|)}
      {2\omega(\vq{1})\left(\omega(\vq{2})+
        \omega(\vq{3})-\omega(\vp{1})\right)
    \,\,2 \omega(\vq{2})\left(\omega(\vq{2})+
        \omega(\vq{3})-\omega(\vq{1})\right)}
     \nonumber \\
  && \delta^{(D-1)}(\vq{2}+\vq{3}-\vp{1})\,\delta^{(D-1)}(\vq{2}+\vq{3}-\vq{1})
  \, (2\pi)^D\,\delta^{(D)} (P-q_1-p_2) \nonumber.
  \eeqa
Note that this expression contains a $D$-dimensional delta function
coming from $\cA^{(0)}$ and two $(D-1)$-dimensional delta functions
coming from 3-momentum conservation of the vertices in the dressing
factor.

\begin{figure}[htb]
 \begin{center}
 \begin{picture}(200,130)(0,50)
 \Photon(190,100)(150,100){3}{5}
 \Line(150,100)(50,150)
 \Line(150,100)(50,50)
 \LongArrow(63,143.5)(60,145)
 \LongArrow(60,55)(63,56.5)
 \GlueArc(100,140)(30,180,307){3.2}{9}
 \GOval(150,100)(12,12)(0){0.9}
 \DashLine(128,140)(128,60){3}
 \LongArrowArcn(100,140)(40,260,225)
 \LongArrow(139,110)(127,116)
 \LongArrow(104,128)(92,134)
 \LongArrow(58,60)(46,54)
 \LongArrow(58,140)(46,146)
 \Text(152,100)[c]{$\cA^{(0)}$}
 \Text(35,150)[c]{$p_1$}
 \Text(35,50)[c]{$p_2$}
 \Text(75,100)[c]{$q_3$}
 \Text(138,120)[c]{$q_1$}
 \Text(102,138)[c]{$q_2$}
 \Text(197,100)[c]{$\alpha$}
 \end{picture}
 \end{center}
 \ccaption{}{Cut diagram for self interaction with 2-particle intermediate
 state. \label{fig:2ptselfint}}
 \end{figure}

We now proceed to perform the integrals over $\vq1$ and $\vq2$,
removing the two $(D-1)$ dimensional delta functions. There is an
important subtlety here. Since the delta functions are $(D-1)$
dimensional, only the spatial part of the 4-vectors is altered.  All
4-vectors in the asymptotic region though must be on-shell and so we
are forced to modify the energy component of these 4-vectors to
preserve this property. Although these modified 4-vectors are 4
component objects they no longer transform as tensors. This is simply
a manifestation of the breaking of Lorentz invariance that occurs in
time-ordered perturbation theory.  To denote such objects we place
curly brackets of the type $\{$ $\}$ around them, i.e. we define
  \beq
  \{p_1-q_3\} \equiv (\omega(\vp1-\vq3), \vp1-\vq3)
  \label{curlydef} \eeq
We then have
  \beqa
  a_{15}^{\{2,0\}} &=&
  -\frac{(-i e)\,g^2}{2}\, T^a_{ik} T^a_{kj}
  \,(2\pi)^D\delta^{(D)} (P-p_1-p_2) \label{eq:a15} \\
  &&  \int \tq3   \left(-g^{\mu\nu} +
  \frac{q_3^{\mu}\,\overline{q}_3^{\nu}+q_3^{\nu}
  \,\overline{q}_3^{\mu}}{(q_3 \overline{q}_3)} \right)
  \Theta(\Delta-|\rho(\vq3,\vp1-\vq3)|)
  \nonumber \\
  &&\frac{\la p_1| \gamma_{\mu} \{\slp1-\slq3\}
  \gamma_{\nu} \slp1 \gamma^{\alpha} |p_2\ra}
  {2\op1 2\opq \rho(\vq3,\vp1-\vq3)^2} \nonumber
  \eeqa
where we defined
 \beq
 \rho(\vk1,\vk2) \equiv \omega(\vk1)+\omega(\vk2)-\omega(\vk1+\vk2)
 \label{rhodef}
 \eeq
This diagram contains infrared singularities coming from the region
where $q_3$ is soft and/or collinear to $p_1$. We discuss its
evaluation  in the Appendix. Multiplying by two to take into account
both of the self-interaction diagrams we get the final result
  \beqa
  2\, a_{15}^{\{2,0\}} &=& 2 \int
  \tq1 \tq2\ g^2\ \cW_{15}^{(2)}(q_{p1},\bq_{p2}; q_{q1},\bq_{q2})
  \times \cA^{(0)}(q_{q1},\bq_{q2}; \gamma(P))
  \label{eq:2_Self_int_Res} \\
  &=& C_F\,\left(\frac{\alpha_{s}}{2\pi}\right)\,
   \left(\frac{\mu^2}{s}\right)^{\epsilon}\Bigg(
   -\frac{1}{\epsilon^2} - \frac{5}{2\epsilon} +
   g_1(\Delta) +\frac{c_R}{2} \nonumber \\
   &&+\ \int\tq3\, \Theta(\Delta-|\rho(\vq3,\vp1-\vq3)|)
   f_1(p_1,p_2,q_3)\Bigg)
  \cA^{(0)}(q_{p1},\bq_{p2}; \gamma(P)) \nonumber
  \eeqa
with
  \beqa
  g_1(\Delta)&=&-\frac{7}{2}-\frac{5}{2}\DoT^2-\frac{7\pi^2}{12}+
  \left[\frac{7}{2}+\DoT+\frac{1}{2}\DoT^2 \right] \lD\nonumber \\
  &+&\ltD+2 \ltpD +4\,\Li\left( \frac{2}{2+\Delta}\right)
  \label{g1def}
  \eeqa
and where $f_1(p_1,p_2,q_3)$ is a function that is free from
singularities when integrated over $\tq3$. The explicit form is given
in \eqn{f1def}. Note that we took care to sum over the physical
polarizations of the gluon only and evaluated the diagram in the
center-of-mass frame $\vp1=-\vp2$.

\subsubsection{The one-gluon exchange terms \label{sec:1_gluon_ext}}

We now look at the one-gluon exchange diagrams. There are two such
diagrams, one for each time ordering of the two vertices. One diagram
is obtained from taking the vertices $V_1, V_8$ of \eqn{vsdef}, the
other from taking the vertices $V_4, V_5$. These diagrams are
symmetric under exchange of all momenta and so we need only calculate
one of them. The diagram shown in Figure~\ref{fig:c6} gives us
 \beqa
 a_{18}^{\{2,0\}} &=& \int
  \tq1 \tq2\ g^2\ \cW^{(2)}_{18}(q_{p1},\bq_{p2}; q_{q1},\bq_{q2})
  \times \cA^{(0)}(q_{q1},\bq_{q2}; \gamma(P)) \label{W18a} \\
  &=&\frac{(-i e)\,g^2}{2} \int d^{D-1}q_1 d^{D-1}q_2
  \frac{d^{D-1}q_3}{(2\pi)^{D-1}}
  \,\,T^a_{ik} T^b_{kj} \nonumber \\
  &&\frac{\delta^{ab}}{2\omega(\vq{3})}
  \left(-g^{\mu \nu}+\frac{q_3^{\mu}\,\overline{q}_3^{\nu}+q_3^{\nu}
  \,\overline{q}_3^{\mu}}{(q_3 \,\overline{q}_3)} \right)
  \la p_1| \gamma_{\mu} \slq{1} \gamma_{\alpha} \slq{2} \gamma_{\nu}|p_2\ra
  \nonumber \\
  &&\frac{\Theta(\Delta-|\oq1+\oq3-\op1|)\,\Theta(\Delta-|\op2+\oq3-\oq2|)}
      {2 \oq1\left(\oq1+\oq3-\op1\right)\,\,2 \oq2\left(\op2+\oq3-\oq2\right)}
      \nonumber \\
  && \delta^{(D-1)}(\vq{1}+\vq{3}-\vp{1})\,\delta^{(D-1)}(\vp{2}+\vq{3}-\vq{2})
  \, (2\pi)^D\,\delta^{(D)} (P-q_1-q_2). \nonumber
 \eeqa

We again integrate over $\vq1$ and $\vq3$ with the delta functions and
introduce the on-shell momenta $\{\slp1-\slq3\}$ and $\{\slp2+\slq3\}$
to obtain
 \beqa
  a_{18}^{\{2,0\}}&=&\frac{(-i e)\,g^2}{2}\,T^a_{ik} T^a_{kj} \,
  \int \frac{d^{D-1}q_3}{(2\pi)^{D-1}}  \,
  (2\pi)^D\,\delta^{(D)} (P-\{p_1-q_3\}-\{p_2+q_3\})\nonumber \\
  &&\frac{1}{2\omega(\vq{3})}
  \left(-g^{\mu \nu}+\frac{q_3^{\mu}\,\overline{q}_3^{\nu}+q_3^{\nu}
  \,\overline{q}_3^{\mu}}{(q_3 \,\overline{q}_3)} \right)
  \la p_1| \gamma_{\mu} \{\slp{1}-\slq3\} \gamma_{\alpha}
  \{\slp2+\slq{3}\} \gamma_{\nu}|p_2\ra
  \nonumber \\
  &&\frac{\Theta(\Delta-|\rho(\vq3,\vp1-\vq3)|)\,
    \Theta(\Delta-|\rho(\vq3,\vp2)|)}
      {2 \omega(\vp1-\vq3)\, \rho(\vq3,\vp1-\vq3)\,
       2 \omega(\vp2+\vq3)\, \rho(\vq3,\vp2)}
      \label{W18b}
 \eeqa
Looking at the denominator $\rho(\vq3,\vp1-\vq3)\,\rho(\vq3,\vp2)$ it
appears that there are collinear singularities for $q_3 \| p_2, \ \
q_3\| p_1$ and a soft singularity $q_3\to 0$. However, the denominator
  \beq
  \left(-g^{\mu \nu}+\frac{q_3^{\mu}\,\overline{q}_3^{\nu}+q_3^{\nu}
  \,\overline{q}_3^{\mu}}{(q_3\,\overline{q}_3)} \right)
  \la p_1| \gamma_{\mu} \{\slp{1}-\slq3\} \gamma_{\alpha}
  \{\slp2+\slq{3}\} \gamma_{\nu}|p_2\ra
  \eeq
vanishes in the collinear regions $q_3 \| p_2$ and $q_3\| p_1$. Thus,
this diagram has only a soft singularity.

 \begin{figure}[htb]
 \begin{center}
 \begin{picture}(200,130)(0,50)
 \Photon(190,100)(150,100){3}{5}
 \Line(150,100)(50,150)
 \Line(150,100)(50,50)
 \LongArrow(63,143.5)(60,145)
 \LongArrow(60,55)(63,56.5)
 \Gluon(80,135)(90,70){3.2}{9}
 \GOval(150,100)(12,12)(0){0.9}
 \DashLine(113,140)(113,60){3}
 \LongArrow(136,112)(125,118)
 \LongArrow(136,88)(125,82)
 \LongArrow(95,94)(93,106)
 \LongArrow(58,60)(46,54)
 \LongArrow(58,140)(46,146)
 \Text(152,100)[c]{$\cA^{(0)}$}
 \Text(35,150)[c]{$p_1$}
 \Text(35,50)[c]{$p_2$}
 \Text(138,120)[c]{$q_1$}
 \Text(138,80)[c]{$q_2$}
 \Text(103,100)[c]{$q_3$}
 \Text(197,100)[c]{$\alpha$}
 \end{picture}
 \end{center}
 \ccaption{}{Cut diagram for the 2-particle cut diagram with one-gluon
  exchange in the asymptotic region.
 \label{fig:c6}}
 \end{figure}

We delegate the explicit evaluation of $a_{18}^{\{2,0\}}$ to the
Appendix. Multiplying by two to account for both one-gluon exchange
diagrams we have the final result
 \beqa
 2\, a_{18}^{\{2,0\}} &=& 2 \int
  \tq1 \tq2\ g^2\ \cW_{18}^{(2)}(q_{p1},\bq_{p2}; q_{q1},\bq_{q2})
  \times \cA^{(0)}(q_{q1},\bq_{q2}; \gamma(P))
  \label{eq:2_1-gluon_exch_Res} \\
  &=& C_F\left(\frac{\alpha_{s}}{2\pi}\right)\,
  \left(\frac{\mu^2}{s}\right)^{\epsilon} \,
   \Bigg(\left( \frac{1}{\epsilon}
    +g_2(\Delta)\right)\cA^{(0)}(q_{p1},\bq_{p2}; \gamma(P)) \nonumber \\
    &-&(-i e)\,\delta_{i j}\,\la p_1|\gamma^{\alpha}|p_2\ra
    (2\pi)^{(D-1)}\delta^{(D-1)} (\vec{P}-\vp1-\vp2) \nonumber \\
    &\times&  \int \tq3\
   \Theta(\Delta-|\rho(\vq3,\vp1-\vq3)|)
   \Theta(\Delta-|\rho(\vq3,\vp2)|) f_2(p_1,p_2,q_3) \Bigg)
  \nonumber
  \eeqa
where again we have not performed the finite $f_2$ integral
analytically and
  \beqa
  g_2(\Delta)=2\log2-2\lD.
  \label{g2def}
  \eeqa
The explicit form of $f_2$ is given in \eqn{f2def}.

\subsubsection{3 Particle Cut Diagram \label{sec:tpcd}}

Let us now turn to the forth part of \eqn{cs2part}. For this term we need the
dressing factor $\cW^{(1)}(q_{p1},\bq_{p2}; q_{q1},\bq_{q2},g_{q3})$. Again we
use the diagrammatic rules of Section~\ref{sec:Feynman_Like_Rules}.

\begin{figure}[ht]
 \begin{center}
 \begin{picture}(200,140)(0,50)
 \Photon(190,100)(150,100){3}{5}
 \Line(150,100)(50,150)
 \Line(150,100)(50,50)
 \LongArrow(63,143.5)(60,145)
 \LongArrow(60,55)(63,56.5)
 \GlueArc(125,138)(40,187.7,300){3.2}{10}
 \GOval(150,100)(12,12)(0){0.9}
 \DashLine(110,140)(110,60){3}
 \LongArrow(136,112)(125,118)
 \LongArrow(136,88)(125,82)
 \LongArrowArcn(125,138)(48,250,220)
 \LongArrow(58,60)(46,54)
 \LongArrow(58,140)(46,146)
 \Text(152,100)[c]{$\cA^{(1)}$}
 \Text(35,150)[c]{$p_1$}
 \Text(35,50)[c]{$p_2$}
 \Text(138,120)[c]{$q_1$}
 \Text(138,80)[c]{$q_2$}
 \Text(90,93)[c]{$q_3$}
 \Text(197,100)[c]{$\alpha$}
 \end{picture}
 \end{center}
 \ccaption{}{Cut diagram for 3-particle intermediate
 state. \label{fig:c4}}
 \end{figure}

There are two possible diagrams as the gluon can be absorbed either by
the quark or antiquark line. The two diagrams are obtained by taking
either the vertex $V_1$ or $V_4$ and they are symmetric under
exchange of momenta. So we need only calculate one of them. For the
diagram shown in Figure~\ref{fig:c4} we get
 \beqa
 a_1^{\{1,1\}} &=& \int
  \tq1 \tq2 \tq3\, g^2\,  \cW^{(1)}_1(q_{p1},\bq_{p2}; q_{q1},\bq_{q2},
   g_{q3})
  \times \cA^{(1)}(q_{q1},\bq_{q2}, g_{q3}; \gamma(P))
  \nonumber \\
  &=&(-i e)\,g^2 \int d^{D-1}q_1 d^{D-1}q_2
  \frac{d^{D-1}q_3}{(2\pi)^{D-1}}\frac{\delta^{ab}}{2\omega(\vq{3})}
  \,T^a_{ik} T^b_{kj}\label{W3ca} \\
  &&\left(-g^{\mu \nu}+\frac{q_3^{\mu}\,\overline{q}_3^{\nu}+q_3^{\nu}
  \,\overline{q}_3^{\mu}}{(q_3\,\overline{q}_3)} \right)
  \frac{\Theta(\Delta-|\oq1+\oq3-\op1|)}{2 \oq1 \left(\oq1+\oq3-\op1
  \right)} \nonumber \\
  &&  \la p_1| \gamma_{\mu} \slq{1}
   \left(\frac{\gamma^{\nu}(\slq1+\slq3)\gamma^{\alpha}}{2(q_1 q_3)}
   -\frac{\gamma^{\alpha}(\slq2+\slq3)\gamma^{\nu}}{2(q_2 q_3)}
  \right)|p_2\ra \nonumber \\
  && \delta^{(D-1)}(\vq{1}+\vq{3}-\vp{1})\,
  \delta^{(D-1)}(\vp{2}-\vq{2})
  (2\pi)^D\,\delta^{(D)} (P-q_1-q_2). \nonumber
 \eeqa

After integration over $\vq1$ and $\vq2$ using the delta functions we
observe that there are collinear singularities $q_3\| p_1$ and soft
singularities $q_3\to 0$. There are, however, no collinear
singularities $q_3\| p_2$. This is expected since the amplitude
$\cA^{(1)}(q_{q1},\bq_{q2}, g_{q3}; \gamma(P))$ has only an integrable
square-root singularity for $q_3\|q_2$ and the dressing factor
$\cW^{(1)}_1(q_{p1},\bq_{p2}; q_{q1},\bq_{q2},g_{q3})$ is regular for
$q_3\|q_2$.

As for the other diagrams we have to multiply by two to take into
account both pairs of diagrams and we get the final result
 \beqa
  2\, a_1^{\{1,1\}} &=& 2 \int
  \tq1 \tq2 \tq3\, g^2\,  \cW_1^{(1)}(q_{p1},\bq_{p2}; q_{q1},\bq_{q2},
   g_{q3})  \label{eq:W1_A1_Amp_Res}\\
  && \hspace*{5cm}
  \times\ \cA^{(1)}(q_{q1},\bq_{q2}, g_{q3}; \gamma(P))
  \nonumber \\
  &=&
  C_F\,\left(\frac{\alpha_{s}}{2\pi}\right)\,
  \left(\frac{\mu^2}{s}\right)^{\epsilon} \,\Bigg(\left(
    \frac{2}{\epsilon^2}
    +\frac{3}{\epsilon}
    +g_3(\Delta)-c_R\right) \
    \cA^{(0)}(q_{p1},\bq_{p2}; \gamma(P)) \nonumber \\
   &+&
    (-i e)\,\delta_{i j}\,\la p_1|\gamma^{\alpha}|p_2\ra
    (2\pi)^{(D-1)}\delta^{(D-1)} (\vec{P}-\vp1-\vp2) \nonumber \\
    &&\hspace*{2cm} \times \int \tq3\
    \Theta(\Delta-|\rho(\vq3, \vp1-\vq3)|) f_3(p_1,p_2,q_3)\Bigg) \nonumber
  \eeqa
where
  \beqa
  g_3(\Delta)&=&7+\DoT^2+\frac{7\pi^2}{6}+\left[-3+2\DoT-\DoT^2\right]\lD
  \nonumber\\
  &-& 2\ltD-4 \ltpD
  -8\, \Li\left(\frac{2}{2+\Delta}\right).
  \label{g3def}
  \eeqa
The function $f_3$ is given in \eqn{f3def} and does not produce any
infrared singularity upon integration over $\tq3$.

\subsubsection{An infrared-finite amplitude}

We have now calculated all terms contributing to the amplitude
$\cA(\{q,\bq\};\gamma)$, \eqn{cs2part}, at next-to-leading order. Using
eqs.~(\ref{cs2part2}, \ref{eq:2_Self_int_Res},
\ref{eq:2_1-gluon_exch_Res}) and (\ref{eq:W1_A1_Amp_Res}) to assemble
the amplitude we get
 \beqa
    \lefteqn{\cA(\{q_{p1}, \bq_{p2}\};\gamma) = } &&
    \label{eq:2_pt_IR_free_amp} \\[5pt]
    &&\hspace*{-0.3cm} 1+C_F\,\left(\frac{\alpha_{s}}{2\pi}\right)
    \Bigg(g_1(\Delta)+g_2(\Delta)+g_3(\Delta)-4+\frac{\pi^2}{12} \Bigg)
    \cA^{(0)}(q_{p1},\bq_{p2}; \gamma(P)) \nonumber \\
    &&\hspace*{-0.3cm}+(-i e)\,\delta_{i j}\,\la p_1|\gamma^{\alpha}|p_2\ra
    (2\pi)^{(D-1)}\delta^{(D-1)} (\vec{P}-\vp1-\vp2)\nonumber \\
    &&\hspace*{-0.3cm}\times \int \tq3\
     \bigg(f_1(p_1,p_2,q_3)\Theta(\Delta-|\rho(\vq3, \vp1-\vq3)|)
   \delta(\sqrt{s}-\op1-\op2)    \nonumber \\
    &&\hspace*{1.2cm} +f_2(p_1,p_2,q_3)\Theta(\Delta-|\rho(\vq3,
    \vp1-\vq3)|)\Theta(\Delta-|\rho(\vq3, \vp2)|)\nonumber \\
    &&\hspace*{1.2cm}
    +f_3(p_1,p_2,q_3)\Theta(\Delta-|\rho(\vq3, \vp1-\vq3)|)\bigg)
   \nonumber
  \eeqa
up to order $\alpha_{s}$ in the coupling. The functions $g_1$, $g_2$,
$g_3$ are given in eqs.~(\ref{g1def}, \ref{g2def}) and (\ref{g3def})
and the functions $f_1$, $f_2$ and $f_3$ are given in
eqs.~(\ref{f1def}, \ref{f2def}) and (\ref{f3def}) respectively.

We see that this result is completely free of infrared
singularities. We are only left with some finite $\Delta$ dependent
terms, $g_i$ and some finite terms, $f_i$ which will in general need
to be numerically integrated. Even though the amplitude $\cA(\{q_{p1},
\bq_{p2}\};\gamma)$ depends on $\Delta$ this dependence will disappear
when we combine the various amplitudes to calculate physical
observables.

\subsection{The amplitude $\cA(\{q(p_1), \bq(p_2),g(p_3)\};\gamma)$
  \label{sec:aqqg}}

We are now going to calculate the amplitude $\cA(\{q_{p1},
\bq_{p2},g_{p3}\};\gamma)$ given in \eqn{cs3part}. There are only two terms to
calculate for this amplitude and there is no integration over the final state
gluon as it is now a real final state particle.

 \begin{figure}[htb]
 \begin{center}
 \begin{picture}(200,150)(0,50)
 \Photon(190,100)(150,100){3}{5}
 \Line(150,100)(50,150)
 \Line(150,100)(50,50)
 \LongArrow(63,143.5)(60,145)
 \LongArrow(60,55)(63,56.5)
 \Gluon(90,130)(50,110){3.2}{7}
 \GOval(150,100)(12,12)(0){0.9}
 \DashLine(113,140)(113,60){3}
 \LongArrow(64,63)(46,54)
 \LongArrow(64,137)(46,146)
 \LongArrow(64,110)(46,101)
 \Text(150,100)[c]{$\cA^{(0)}$}
 \Text(35,150)[c]{$p_1$}
 \Text(35,50)[c]{$p_2$}
 \Text(35,100)[c]{$p_3$}
 \Text(197,100)[c]{$\alpha$}
 \end{picture}
 \end{center}
 \ccaption{}{Cut diagram for the 3-particle asymptotic region with
 a 2-particle  intermediate state.
   \label{fig:3pt_w1_a0}}
 \end{figure}

Again we calculate these terms using the diagrammatic rules from
Section~\ref{sec:Feynman_Like_Rules}. Let us start with the diagrams
where the gluon is emitted in the dressing factor.
Figure~\ref{fig:3pt_w1_a0} shows one of the two possible diagrams, the
other is exactly the same but with all momenta interchanged. So for
both diagrams we have
  \beqa
  \lefteqn{\int\,\tq1 \tq2\ g\,
  \cW^{(1)}_{1}(q_{p1},\bq_{p2}, g_{p3}; q_{q1},\bq_{q2})
  \times\cA^{(0)}(q_{q1},\bq_{q2}; \gamma(P))} &&\nonumber \\
  &=&(-i
  e)g\,T_{ij}^a\,(2\pi)^{D}\delta^{(D-1)}(\vec{P}-\vp1-\vp2-\vp3)\,
  \la p_1| \\[5pt]
  && \hspace*{-1cm} \Bigg( -\,\frac{\,\,\sleps_{p_3}\{\slp1+\slp3\}
  \gamma^{\alpha}}{2\omega(\vp1+\vp3)r_1} \Theta(\Delta-|r_1|)
  \delta(\sqrt{s}-\op1-\op2-\op3+r_1) \nonumber \\
  &+&\!
   \frac{\gamma^{\alpha}\{\slp2+\slp3\}\,\sleps_{p_3}}{2\omega(\vp2+\vp3)
   r_2}
   \Theta(\Delta-|r_2|)\delta(\sqrt{s}-\op1-\op2-\op3+r_2)\Bigg)|
   p_2\ra
  \nonumber
  \eeqa
where we used the notation
  \beqa
  r_1 &\equiv& \rho(\vp1,\vp3) = \op1+\op3-\omega(\vp1+\vp3)
  \label{ridef} \\
  r_2 &\equiv& \rho(\vp2,\vp3)=\op2+\op3-\omega(\vp2+\vp3),
  \nonumber \eeqa
with $\rho$ defined in \eqn{rhodef}.

\begin{figure}[ht]
 \begin{center}
 \begin{picture}(200,150)(0,50)
 \Photon(190,100)(150,100){3}{5}
 \Line(150,100)(50,150)
 \Line(150,100)(50,50)
 \LongArrow(63,143.5)(60,145)
 \LongArrow(60,55)(63,56.5)
 \Gluon(140,100)(50,100){3.2}{12}
 \GOval(150,100)(12,12)(0){0.9}
 \DashLine(110,140)(110,60){3}
 \LongArrow(64,63)(46,54)
 \LongArrow(64,137)(46,146)
 \LongArrow(64,93)(46,93)
 \Text(150,100)[c]{$\cA^{(1)}$}
 \Text(35,150)[c]{$p_1$}
 \Text(35,50)[c]{$p_2$}
 \Text(35,93)[c]{$p_3$}
 \Text(197,100)[c]{$\alpha$}
 \end{picture}
 \end{center}
 \ccaption{}{Cut diagram for 3-particle asymptotic region with a 3-particle
 intermediate
 state.
   \label{fig:3pt_w0_a1}}
 \end{figure}

The second contribution is just the usual
$\cA^{(1)}(q_{p1},\bq_{p2},g_{p3}; \gamma(P))$ amplitude. The three
external particles of this amplitude do not interact in the asymptotic
region and so we simply have the diagram as shown in
Figure~\ref{fig:3pt_w0_a1}, this gives

  \beqa
  \int\, &\hspace*{-0.3cm}\tq1 &\hspace*{-0.3cm} \tq2\ g\,
  \cW^{(0)}(q_{p1},\bq_{p2},g_{p3}; q_{q1},\bq_{q2},g_{q3})\times
  \cA^{(1)}(q_{q1},\bq_{q2},g_{q3}; \gamma(P)) \nonumber\\
  &=&(-ie)g\,T_{ij}^a\,\la p_1|
  \Bigg(\frac{\,\,\sleps_{p_3}(\slp1+\slp3)\gamma^{\alpha}}{2(p_1 p_3)}
  -\frac{\gamma^{\alpha}(\slp2+\slp3)
  \,\,\sleps_{p_3}}{2(p_2 p_3)}\Bigg)|p_2 \ra \nonumber \\
  &&(2\pi)^{D}\delta^{(D)}(P-p_1-p_2-p_3).
  \eeqa

We now assemble \eqn{cs3part} to find
 \beqa
 \cA(\{q_{p1}, \bq_{p2},g_{p3}\};\gamma)&=&
 (-i e)g\,T_{ij}^a\,\la p_1|\Bigg(\nonumber \\
  &&\hspace*{-4cm}-\frac{\,\,\sleps_{p_3}\{\slp1+\slp3\}
  \gamma^{\alpha}}{2\omega(\vp1+\vp3)r_1}
   \Theta(\Delta-|r_1|)\delta(\sqrt{s}-\op1-\op2-\op3+r_1) \nonumber \\
  &&\hspace*{-4cm}+\frac{\,\,\sleps_{p_3}(\slp1+\slp3)
  \gamma^{\alpha}}{2(p_1 p_3)}\delta(\sqrt{s}-\op1-\op2-\op3) \nonumber \\
  &&\hspace*{-4cm}+\frac{\gamma^{\alpha}\{\slp2+\slp3\}
  \,\,\sleps_{p_3}}{2\omega(\vp2+\vp3)r_2}
   \Theta(\Delta-|r_2|)\delta(\sqrt{s}-\op1-\op2-\op3+r_2) \nonumber \\
  &&\hspace*{-4cm}-\frac{\gamma^{\alpha}(\slp2+\slp3)
  \,\,\sleps_{p_3}}{2(p_2 p_3)}\delta(\sqrt{s}-\op1-\op2-\op3)
 \Bigg)| p_2\ra \nonumber \\
  &&\hspace*{-4cm}(2\pi)^{D}\delta^{(D-1)}(\vec{P}-\vp1-\vp2-\vp3).
 \label{eq:3_particle_final_state}
 \eeqa
This amplitude splits up into two pairs. The first (last) two terms
are due to the gluon being emitted from the leg $p_1$ ($p_2$).
Looking at the first two terms  shows that for $r_1>\Delta$ the
contribution from 
the asymptotic region disappears.  We are then left with the normal
amplitude, $\cA(q_{p1}, \bq_{p2},g_{p3};\gamma)$.  For $r_1<\Delta$
the term from the asymptotic region does contribute and will cancel
any potential infrared singularities. We can see this by taking the
limit $\Delta\rightarrow 0$, we have 
 \beqa
 \omega(\vp1+\vp3)r_1&\rightarrow&(p_1 p_3), \nonumber \\
 \{ \slp1+\slp3\}&\rightarrow&( \slp1+\slp3), \nonumber \\
 \delta(\sqrt{s}-\op1-\op2-\op3+r_1)&\rightarrow&
 \delta(\sqrt{s}-\op1-\op2-\op3). \nonumber
 \eeqa
With these we can see that the terms from the asymptotic region
approach those of the normal amplitude in the soft and collinear
limits, but with the opposite sign.  So the two terms will cancel
in the $\Delta\rightarrow0$ limit, leaving us with an amplitude
that is infrared finite when integrated over the phase space.

\subsection{Calculation of the total cross section}

In the previous sections we computed the two infrared finite
amplitudes that contribute to the process $\gamma^*(P)\to 2$ jets at
next-to-leading order. In this section we would like to check our
results by computing the total cross section, starting from the
infrared finite amplitudes,
\eqns{eq:2_pt_IR_free_amp}{eq:3_particle_final_state}. Of course, we
have to recover the well known result, \eqn{totXS}.

Let us stress that the idea of our approach is to compute the
amplitudes numerically and perform the phase-space integration also
numerically. It is for the sole purpose of checking our results and
facilitating the comparison with \eqn{totXS} that in this section we
compute the total cross section analytically.

Usually a non zero value of $\Delta$ would be chosen for a numerical
calculation and we would expect all $\Delta$ dependence to cancel
between the contributions of the two amplitudes (squared) to the cross
section. Here though to simplify the analytical calculation we will
take the limit $\Delta\rightarrow0$. In this limit even the
infrared-finite amplitudes are proportional to a four-dimensional
delta function and we can use the standard procedure to obtain the
total cross section from the amplitudes. However, since the amplitudes
are singular for $\Delta\to 0$ we must be careful in taking this limit
and leave it until the end of the calculation. The $f_n$ finite terms
of \eqn{eq:2_pt_IR_free_amp} which we would usually have to calculate
numerically will all go to zero in this limit. This simplification
occurs because the region of integration shrinks to zero as
$\Delta\rightarrow0$ and as these terms are finite they can no longer
give a contribution.

We now use our infrared finite amplitudes \eqn{cs2part} and
\eqn{cs3part} instead of \eqn{sm0} and \eqn{sm1} and square them in
the usual way  to obtain
 \beqa
 \sigma=\sigma_{\{q\overline{q}\}}+\sigma_{\{q\overline{q}g\}},
 \label{eq:Total_Cross_section} \eeqa where \beqa
 \sigma_{\{q\overline{q}\}}= \int {\rm d}\Phi_2
 \left|\cA(\{q_{p1},\overline{q}_{p2}\};\gamma)\right|^2,
 \label{eq:qqb_Cross_section} \\ \sigma_{\{q\overline{q}g\}}=
 \int {\rm d}\Phi_3
 \left|\cA(\{q_{p1},\overline{q}_{p2},g_{p3}\};\gamma)\right|^2.
 \label{eq:qqbg_Cross_section}
 \eeqa
Here we integrate \eqn{eq:qqb_Cross_section} over the two particle
phase space and \eqn{eq:qqbg_Cross_section} over the three particle
phase space.

First we rewrite the three-particle final state amplitude,
\eqn{eq:3_particle_final_state}, in a more convenient form
 \beqa
 \cA(\{q_{p1}, \bq_{p2},g_{p3}\};\gamma)&=&(-i e)g\,T_{ij}^a\,\la p_1|
 \Bigg(\nonumber \\
  &&\hspace*{-4cm}\left(-\frac{\,\,\sleps_{p_3}\{\slp1+\slp3\}
  \gamma^{\alpha}}{2\omega(\vp1+\vp3)r_1}\delta(E+r_1)
   +\frac{\,\,\sleps_{p_3}(\slp1+\slp3)\gamma^{\alpha}}{2(p_1 p_3)}\delta(E)
   \right)\Theta(\Delta-|r_1|) \nonumber \\
  &&\hspace*{-4cm}+\frac{\,\,\sleps_{p_3}(\slp1+\slp3)
  \gamma^{\alpha}}{2(p_1 p_3)}
  \delta(E)\Theta(|r_1|-\Delta) \nonumber \\
  &&\hspace*{-4cm}+\left(\frac{\gamma^{\alpha}\{\slp2+\slp3\}
  \,\,\sleps_{p_3}}{2\omega(\vp2+\vp3)r_2}\delta(E+r_2)
  -\frac{\gamma^{\alpha}(\slp2+\slp3)\,\,\sleps_{p_3}}{2(p_2 p_3)}
  \delta(E)\right)
   \Theta(\Delta-|r_2|)\nonumber \\
  &&\hspace*{-4cm}-\frac{\gamma^{\alpha}(\slp2+\slp3)\,\,
  \sleps_{p_3}}{2(p_2 p_3)}
  \delta(E)\Theta(|r_2|-\Delta)\Bigg)
  | p_2\ra \nonumber \\
  &&\hspace*{-4cm}(2\pi)^{D}\delta^{(D-1)}(\vec{P}-\vp1-\vp2-\vp3),
  \label{eq:qqbg_Cross_section_simp}
 \eeqa
where $E=\sqrt{s}-\op1-\op2-\op3$. Taking \eqn{eq:qqbg_Cross_section_simp}
we then square it in the usual way and sum over the gluon
polarizations using
  \beqa
 \sum  \eps^{\mu}_{p_3}\eps^{\nu}_{p_3}=-g^{\mu\nu}+
                     \frac{p^{\mu}_{3}\overline{p}^{\nu}_{3}+
                     p^{\nu}_3\overline{p}^{\mu}_{3}}{
                     (p_{3}\overline{p}_{3})} \nonumber
  \eeqa
where $\overline{p}_{3}=(\op{3},-\vp{3})$. This is because the
amplitude is no longer gauge invariant as we are using dressed
states. At this point we drop any terms multiplied by
$\Theta(\Delta-|r_1|)$ or $\Theta(\Delta-|r_2|)$. These terms are finite
and therefore can be shown to go to zero in the $\Delta\rightarrow0$
limit after we have performed the three particle phase space integral
in a similar way to the $f_n$ terms.

After integrating one of the phase space integrals using the delta
function we are left with
 \beqa
 &&\hspace*{-0.8cm}
 \left|\cA(\{q_{p1}, \bq_{p2},g_{p3}\};\gamma)\right|^2=
 4\,C_F\,(2\pi)^{3-2D}
 \int \frac{d\Omega_{D-1}}{2^{D-1}}d\Omega_{D-2} \\
 &&\Bigg(\int^{\frac{\Delta}{2}}_{0}dy_{13}
 \int^{0}_{1-y_{13}}dy_{23}
 \frac{2y_{23}-y_{13}\left(y_{13}+y_{23}\right)^2}
      {y_{13}\left(y_{13}+y_{23}\right)^2}
      \nonumber \\
 &&-\int^{1-\frac{\Delta}{2}}_{0}dy_{13}
 \int^{\frac{\Delta}{2}}_{1-y_{13}}
 dy_{23}\,
 \frac{y_{23}^3+y_{13}^2y_{23}+2y_{13}
 \left(y_{23}-1\right)^2}
      {y_{23}\left(y_{13}+y_{23}\right)^2} \nonumber \\
 &&+\int^{1-\frac{\Delta}{2}}_{\frac{\Delta}{2}}dy_{13}
 \int^{\frac{\Delta}{2}}_{1-y_{13}}dy_{23} \Bigg(2
 -\frac{2-y_{23}}{y_{13}}-\frac{2-y_{13}}{y_{23}}
 +\frac{4}{\left(y_{13}+y_{23}\right)^2}\Bigg)
 \Bigg)\nonumber
 \eeqa
where we defined
 \beq
 y_{ij} \equiv \frac{2(p_i p_j)}{\xi_{p_1}^2}.
 \eeq
We perform the final two integrals and then prematurely take the
$\Delta\rightarrow0$ limit everywhere except in the $\log (\Delta)$
terms, as these diverge in this limit. The $\log(\Delta)$ terms will
cancel later in the final result. This then leaves
 \beqa
 \frac{1}{\sigma_0}\sigma_{\{q\overline{q}g\}}=
 C_F\left(\frac{\alpha_{s}}{\pi}\right)\left(\frac{5}{4}-\log4
 +\frac{3}{2}\log\left(\frac{\Delta}{2}\right)+
 \log^2\left(\frac{\Delta}{2}\right)\right)
 \label{eq:qqbg_cross_section}
 \eeqa
where $\sigma_0$ is the total Born cross section as given in \eqn{smsg0}.

Now we calculate $\left|\cA(\{q_{p1},
\bq_{p2}\};\gamma)\right|^2$. Again we take the $\Delta\rightarrow0$
limit early except for the $\log(\Delta)$ pieces of the $g_n$ terms in
the finite part of \eqn{eq:2_pt_IR_free_amp}. As stated before the
$f_n(p_1,p_2,q_3)$ terms go to zero and so we have,
 \beqa
 &&\hspace*{-0.8cm}
 \left|\cA(\{q_{p1}, \bq_{p2}\};\gamma)\right|^2=
 \left|\cA^{(0)}(q_{p1},\bq_{p2}; \gamma(P))\right|^2 \\
 &\times&\Bigg(1+C_F\left(\frac{\alpha_{s}}{2\pi}\right)
 \left(-\frac{1}{2}+\log4-\frac{3}{2}
 \log\left(\frac{\Delta}{2}\right)
 -\log^2\left(\frac{\Delta}{2}\right)\right)\Bigg)^2 \nonumber
 \eeqa
After integrating over the two particle phase space we get
 \beq
 \frac{\sigma_{\{q\overline{q}\}}}{\sigma_{0}} =
 \left(1+C_F\,\frac{\alpha_{s}}{\pi}
 \left(-\frac{1}{2}+\log4-
 \frac{3}{2}\log\left(\frac{\Delta}{2}\right)-
 \log^2\left(\frac{\Delta}{2}\right)
    \right)+\cO(\alpha_{s}^2)\right)
 \label{eq:qqb_cross_section}
 \eeq
Putting \eqn{eq:qqbg_cross_section} and \eqn{eq:qqb_cross_section} together
gives finally
 \beq
 \frac{\sigma}{\sigma_{0}} =
 \left(1+\left(\frac{\alpha_{s}}{\pi}\right)\frac{3}{4}\,C_F
  +\cO(\alpha_{s}^2)\right).
 \eeq
We have recovered the well known result for the total
$\gamma\rightarrow q\overline{q}$ cross section and all the $\Delta$
dependence of the amplitudes has disappeared including the
$\log(\Delta)$ terms, justifying our taking of the $\Delta
\rightarrow0$ limit early.

\section{Summary and outlook \label{summary}}

We have presented a method on how to construct infrared finite
amplitudes and applied it to the case of $e^+e^-\to2$ jets at
next-to-leading order in the strong coupling. The idea is to separate
from the Hamiltonian a part that describes the asymptotic
dynamics. This asymptotic Hamiltonian is then used to asymptotically
evolve the usual states of the Fock space. In this way we construct
dressed states, \eqns{iasdef}{fasdef}, such that the transition
amplitudes between these states are free from infrared singularities.

Contrary to most of the previous work done in this field we are not so much
interested in obtaining all-order resumed results taking into account soft
emission of an arbitrary number of gauge bosons from external partons. Our aim
is to construct dressed states explicitly order-by order in perturbation
theory and use them to do explicit calculations. In this paper we have done
this for a particularly simple final state up to next-to-leading order. In the
future we would like to expand this to more complicated external states and
higher orders.

The reason that we cannot obtain all-order results is that we include the
collinear singularities as well. In non-abelian theories these singularities
cannot be avoided. The additional complications due to the collinear
singularities make it impossible to obtain exact solutions to the asymptotic
dynamics. Collinear singularities have been considered
previously~\cite{Havemann:1985ra, DelDuca:jt, Contopanagos:1991yb} but to the
best of our knowledge the amplitudes presented in this work are the first
infrared-finite amplitudes for a realistic scattering process in QCD.

As for the standard approach, physical cross sections obtain in
general contributions from more than one partonic process. However, in
our case all these contributions are separately finite. They depend on
a parameter, $\Delta$, that determines the precise split of the
Hamiltonian into an asymptotic Hamiltonian and the remainder. The
result for any physical quantity is independent of this parameter as
long as it is smaller than any experimental resolution.  For any
finite value of $\Delta$ the amplitude contains a part that is not
proportional to an energy conserving delta function which represents
the spread of the initial wave packet due to the asymptotic evolution.

For any physical cross section at any order in perturbation theory we
will get the same answer using the standard cross-section method or
infrared-finite amplitudes. Thus, one might wonder what has been
gained using this approach. Apart from the conceptional benefit that
the $S$-matrix between dressed states is well defined there are also
practical advantages. First of all, the avoidance of infrared
singularities facilitates the use of numerical methods. This might not
be apparent in the approach we have taken. In fact, using \eqn{facS}
to split the infrared finite amplitudes into separately divergent
factors still requires us to use an infrared regulator (dimensional
regularization in our case) and revert to analytical calculations.
However, since the final amplitude is infrared finite it is feasible
to compute it directly in a numerical way, avoiding the split into
separately divergent pieces. Once the amplitudes have been obtained,
the integration over the phase space is trivial and no sophisticated
method is needed. This also opens up the possibility of combining
fixed-order calculations directly with a parton shower approach.

Needless to say that the explicit example we considered, $e^+e^-\to 2$
jets has many simplifying features. To start with, the non-abelian
nature of QCD does not really enter. Secondly, we only considered the
amplitudes at next-to-leading order. Furthermore, the initial state
does not interact strongly.

The last point simply results in the fact that there is no need to dress the
initial state. While this is a simplification concerning the amount of
computations to be performed, there is no conceptual problem associated with
more complicated initial states. If the initial state contains hadrons a
physical cross section is obtained by folding the partonic cross section with
parton densities. In the conventional approach these parton densities are
associated with the probability of finding a certain partonic state within a
hadron. In our case, we would have to use modified parton densities that are
related to the probability of finding a certain dressed state within a
hadron. Thus the global analyzes of extracting the parton densities would have
to be modified and repeated.

The fact that the non-abelian nature of QCD does not really show up in
the explicit example we considered results in a particularly simple
asymptotic Hamiltonian. In fact, the asymptotic Hamiltonian we use
involves only quark-gluon interactions and is basically the same that
was used many times previously~\cite{Contopanagos:1991yb}. Again, this
results in a technical simplification of the computation and
facilitates the explicit construction of the asymptotic
Hamiltonian. In more complicated examples the full non-abelian
structure of the asymptotic Hamiltonian will enter the problem and its
construction will be much more involved. However, the only crucial
feature is that the asymptotic Hamiltonian reproduces the full
asymptotic dynamics, i.e.  it has to reproduce the soft and collinear
behavior of the full theory. There are no further requirements and the
construction of dressed states presented in this paper can be taken
over directly. However, it is clear that the construction used so far
is rather cumbersome. In order to exploit the advantage of the
infrared finiteness a systematic numerical approach should be
developed. This will become particularly important if the method is to
be extended beyond next-to-leading order.

\section*{Appendix}

In this Appendix we give some details concerning the evaluation of the
diagrams mentioned in Section~\ref{sec:aqq}.

We consider first  with the self-interaction term $a_{15}^{\{2,0\}}$ of
Section~\ref{sec:si}. We start from \eqn{eq:a15},
substitute $\slp1-\slq3+\slr{}$ for $\{\slp1-\slq3\}$, where
$r=(r_0,\vec0)$ with $r_0=\rho(\vq3,\vp1-\vq3)$ and
then expand the numerator to obtain
  \beqa
  a_{15}^{\{2,0\}}&=&\frac{(-i e)\,g^2}{4}\,T^a_{ik} T^a_{kj}\,
  (2\pi)^D\delta^{(D)} (P-p_1-p_2)
  \int \tq3 \,\,\Theta(\Delta-|r_0|) \label{W15b} \\
  &&\Bigg((D-2)((p_1 q_3)-(p_1 r)) -
  \frac{4(p_1 q_3)(p_1 \overline{q}_3)}{(q_3\overline{q}_3)} -
  \frac{2r_0(p_1 q_3)(p_2 q_3)}{\op1(q_3\overline{q}_3)}
  \Bigg)\, \nonumber \\
  &&\frac{\la p_1|\gamma^{\alpha}|p_2\ra}{\op1\,\opq\, r_0^2}
  \nonumber
  \eeqa
This expression contains infrared singularities coming from the region
where $q_3$ is soft and/or collinear to $p_1$. In order to evaluate
the expression, \eqn{W15b} we choose to parameterize the momenta in
the center-of-mass frame. The momenta are all on-shell and are defined
as
  \beqa
  P&=&\sqrt{s}(1,\mathbf{0},0),\label{mompar} \\
  p_1&=& \frac{\xi_{p_1}}{2}(1,\mathbf{0},1),\nonumber \\
  p_2&=& \frac{\xi_{p_1}}{2}(1,\mathbf{0},-1),\nonumber \\
  q_3&=& \frac{\xi_{p_1}}{2}\,z\,(1,\sqrt{1-y^2}\, \mathbf{e}_T,y),
 \nonumber \\
  \{p_1-q_3\}&=& \frac{\xi_{p_1}}{2}(\sqrt{1-2 z y+z^2},-z \sqrt{1-y^2}\,
  \mathbf{e}_T,1-z y),
  \nonumber
  \eeqa
where $\mathbf{0}$ is the null vector in a $(2-2\epsilon)$-dimensional
space, $\mathbf{e}_T$ is a unit vector in the
$(2-2\epsilon)$-dimensional transverse momentum space and we have
$0\leq z \leq \infty$, $-1\leq y\leq 1$. The singular limits are then
given by the limits $z\rightarrow0$ for soft singularities,
$y\rightarrow1$ for $q_3\| p_1$ singularities and $y\rightarrow-1$ for
$q_3\| p_2$ singularities. As the asymptotic region does not conserve
energy we find that the upper limit of $z$ goes to $\infty$. This
would suggest the possibility of UV singularities in the asymptotic
regions.  However we will see that the $\Theta$ function will restrict
this upper limit to a finite value, removing the need to renormalize
these regions.  The integral measure is given by
  \beqa
  \lefteqn{\int \tq3 \,\,\Theta(\Delta-r_0)\rightarrow} &&
  \label{pspar} \\
  && \left(\frac{\mu^2}{s}\right)^{\epsilon} \frac{1}{2(2\pi)^{3-2\epsilon}}
  \,\, \frac{\xi_{p_1}^2}{4} \, z^{1-2\epsilon}(1-y)^{-\epsilon}\,
  (1+y)^{-\epsilon}\,
  dy \, dz \, d\Omega_{(2-2\epsilon)} \nonumber
  \eeqa
where we have three separate integration regions for the $z$ and $y$ integrals,
   \beqa
  0\leq z\leq\frac{2}{\Delta}\frac{(2+\Delta)}{(1-y+\Delta)}
  &\textrm{with}&-1\leq y\leq\frac{2-\Delta^2}{2}, \nonumber \\
  0\leq z\leq1&\textrm{with}&\frac{2-\Delta^2}{2}\leq y\leq1,\nonumber \\
  1\leq z\leq\frac{2}{\Delta}\frac{(2+\Delta)}{(1-y+\Delta)}
  &\textrm{with}&\frac{2-\Delta^2}{2}\leq y\leq1.\nonumber
  \eeqa
The infrared singularities are in the first two regions whereas the
last region will  give a finite contribution. The remaining angular
integral is given by
  \beqa
  \int d\Omega_{(2-2\epsilon)}=\frac{2 \pi^{1-\epsilon}}{\Gamma(1-\epsilon)}.
  \label{eq:angular_int}
  \eeqa
We now turn back to \eqn{W15b} and notice that the infrared
singularities $q_3$ soft and/or collinear to $p_1$ come from the
region $z=0$ and $y=1$ but not $y=-1$. We use the subtraction method
to isolate these singularities and evaluate them analytically. Writing
the integrand schematically as a function $F(z,y)$ we write
  \beqa
  F(z,y)&=&\left(F(0,y)+F(z,1)-F(0,1)\right)\label{Fsplit} \\
  &+&\left(F(z,y)-F(0,y)-F(z,1)+F(0,1)\right). \nonumber
  \eeqa
The first term contains all the divergent pieces whereas the second
term will give a finite contribution upon integration over
$\tq3$. Applying this method to \eqn{W15b} we obtain
  \beqa
  a_{15}^{\{2,0\}} &=& (-i e) \,g^2\,T^a_{ik} T^a_{kj}\,
  \left(\frac{\mu^2}{s}\right)^{\epsilon}\,
  \frac{1}{2(2\pi)^{3-2\epsilon}}\,
  (2\pi)^D\delta^{(D)} (\vec{P}-\vp1-\vp2)
  \label{W15e} \\
  && \la p_1|\gamma^{\alpha}|p_2\ra
  \int d\Omega_{(2-2\epsilon)} \, dy \, dz
  \,z^{1-2\epsilon}\,(1-y)^{-\epsilon}
  \, (1+y)^{-\epsilon}  \nonumber \\
  && \qquad \qquad
  \times\ \left(\frac{(2-D)}{4(1-y)}+ \frac{(2z-1-y)}{2(1-y)z^2} +
  f_1(p_1,p_2,q_3)\right) \nonumber
  \eeqa
where
  \beqa
  \lefteqn{f_1(p_1,p_2,q_3) =} && \label{f1def} \\
  && \frac{\op1^2}{2\opq\, r_0}\Bigg(1 -
  \frac{(p_1q_3) }{\op1 \, r_0}-
  \frac{(p_1 q_3)(p_2 q_3)}{\op1 r_0(q_3 \overline{q}_3)}
  \left(2-\frac{r_0}{\op1}\right)
  \Bigg) \nonumber \\
  &&\hspace*{0.3cm} -\ \frac{\op1^2}{2(p_1 q_3)}\Bigg(
  -\frac{\oq3}{\op1}-\frac{(p_2 q_3)}{\oq3^2}+2
  \Bigg).
  \eeqa
Integrating the singular terms and expanding around $\epsilon=0$ we
obtain \eqn{eq:2_Self_int_Res}. Note that the $D$ in the first term of
\eqn{W15e} arises from the $\gamma$-matrix algebra. Thus we write it
as $D=4-2\eps+c_R 2 \eps$ to obtain the expressions in conventional
dimensional regularization ($c_R=0$) and in dimensional
reduction ($c_R=1$).

Let us now turn to the evaluation of $a_{18}^{\{2,0\}}$ needed in
Section~\ref{sec:1_gluon_ext}. We start with the expression \eqn{W18b}
and proceed in the same way as for $a_{15}^{\{2,0\}}$.  We introduce
the on-shell momenta $\{\slp1-\slq3\} = \slp1-\slq3+\slr{}$ and
$\{\slp2+\slq3\}= {\slp2}+\slq3-\slr{}'$, where $r'=(r_0',\vec0)$ with
$r_0'=\rho(\vq3,\vp2) = \omega(\vq3)+\omega(\vp2)-\omega(\vp2+\vq3)$.
In order to proceed we subtract the soft singularity in \eqn{W18b} and
add it back to produce an integrand that results in a non-singular
term. In the soft limit the $D$-dimensional delta function becomes the
usual $\delta^{(D)} (P-p_1-p_2)$ which can be pulled out from the
integral.

We use the same momentum parametrization as for the self-interacting case, but
because of the extra $\Theta$ function the integration ranges change to
  \beqa
  0\leq z\leq\frac{2}{\Delta}\frac{(2+\Delta)}{(1-y+\Delta)}
  &\textrm{with}&-1\leq y\leq\frac{\Delta}{2}, \nonumber \\
  0\leq z\leq-\frac{2}{\Delta}\frac{(2-\Delta)}{(1+y-\Delta)}
  &\textrm{with}&\frac{\Delta}{2}\leq y\leq1 \nonumber
  \eeqa
The remaining angular integral is as given in \eqn{eq:angular_int}.

Using this momentum parametrization and expanding around  the soft region gives
  \beqa
  a_{18}^{\{2,0\}} &=& -T^a_{ik} T^a_{kj}\,
  \left(\frac{\mu^2}{s}\right)^{\epsilon} (-i e)\,g^2\,
  \frac{1}{2(2\pi)^{3-2\epsilon}}\,
  (2\pi)^D\delta^{(D-1)} (\vec{P}-\vp1-\vp2) \nonumber \\
  &&  \la p_1| \gamma^{\alpha}|p_2\ra
  \int d\Omega_{(2-2\epsilon)}
  \, dy \, dz \,z^{1-2\epsilon}\,(1-y)^{-\epsilon}
  \, (1+y)^{-\epsilon} \nonumber \\
  &&\qquad \qquad \times \
  \left(\frac{1}{2 z^2}\delta (\sqrt{s}-\op1-\op2)+
  f_2(p_1,p_2,q_3)\right)
  \eeqa
with
 \beqa
 f_2(p_1,p_2,q_3)&=&
 \frac{\op1^2}{2\opq\, r_0\,\omega(\vp2+\vq3)\,r_0'}
 \Bigg((p_1 p_2)+(p_1 q_3)-(p_2 q_3) \nonumber \\
 &+&\op1 r_0 - \op1 r'_0-\frac{r_0 r'_0}{2}+\frac{(p_1 q_3)r_0}{2\op1}
 +\frac{(p_2 q_3)r'_0}{2\op1} \nonumber \\
 &-&\frac{(p_1 q_3)(p_2 q_3)}{2\op1^2}
 -\frac{1}{2(q_3 \overline{q}_3)}
 \Bigg(\frac{(p_1 q_3)^3}{\op1^2}\left(1+\frac{r_0}{2\op1}\right) \nonumber \\
 &-&\frac{(p_2 q_3)^3}{\op1^2}\left(1-\frac{r'_0}{2\op1}\right) +
   \left(2+\frac{r_0}{\op1}-\frac{r'_0}{\op1}-
   \frac{r_0 r'_0}{\op1^2}\right) \nonumber \\
 &&\left((p_1 q_3)^2+(p_2 q_3)^2\right)\Bigg)\Bigg)
 \delta (\sqrt{s}-\op1-\op2-r_0+r_0') \nonumber \\
 &-&\frac{\op1^2}{2\oq3^2} \delta (\sqrt{s}-\op1-\op2).
 \label{f2def}
 \eeqa
Upon performing the integration of the singular terms explicitly and expanding
in $\epsilon$ we get \eqn{eq:2_1-gluon_exch_Res}. In this case the
expression is the same in conventional dimensional regularization and
dimensional reduction.

Finally we turn to the evaluation of $a_1^{\{1,1\}}$ needed in
Section~\ref{sec:tpcd}, proceeding as in the previous cases. We subtract the
soft and collinear singular parts and integrate them analytically. In both
limits the $D$-dimensional delta function takes its usual form
$\delta^{(D)}(P-p_1-p_2)$. Thus, the delta function is independent of the
integration variables and can be taken outside the integral, as in the
one-gluon exchange terms.

We can use the same momentum parametrization and integration regions as the
self-interacting case as we have the same $\Theta$ function in both cases.
Taking the $z\rightarrow0$ and $y\rightarrow1$ limits of the above terms we
obtain
  \beqa
  a_1^{\{1,1\}} &=& (-i e)\,g^2\, T^a_{ik} T^a_{kj}
  \,\left(\frac{\mu^2}{s}\right)^{\epsilon}\,
  \frac{1}{2(2\pi)^{3-2\epsilon}}\,
  (2\pi)^D\delta^{(D-1)} (\vec{P}-\vp1-\vp2)
   \nonumber \\
  &&\la p_1|\gamma^{\alpha}|p_2\ra \int d\Omega_{(2-2\epsilon)}
  \, dy \, dz \,z^{1-2\epsilon}\,(1-y)^{-\epsilon}
  \, (1+y)^{-\epsilon}  \\
  && \times \ \Bigg(\frac{4-4z+(D-2)z^2}{2 (1-y) z^2}
  \delta(\sqrt{s}-\op1-\op2)
  +f_3(p_1,p_2,q_3)\Bigg) \nonumber
  \eeqa
where
  \beqa
  f_3(p_1,p_2,q_3)&\hspace*{-0.3cm}=&\hspace*{-0.3cm}
  \frac{\op1^2}{2\opq r_0 (\{p_1-q_3\}q_3)}\Bigg(r^2_0
  +2\,r_0\,\op1 \nonumber \\
  &\hspace*{-0.3cm}-&\hspace*{-0.3cm}
  (p_1 q_3)\left(2+\frac{r_0}{\op1}\right)
  +\frac{(p_1 q_3)(p_2 q_3)}{(q_3 \overline{q}_3)}
  \left(2+\frac{2r_0}{\op1}+\frac{r^2_0}{2\op1^2}\right)\Bigg) \nonumber \\
  &\hspace*{-0.3cm}+&\hspace*{-0.3cm}
  \frac{\op1^2}{2\opq r_0 (p_2 q_3)}\Bigg(2(p_1 p_2)+(p_1 q_3)\left(2
  +\frac{r_0}{\op1}-\frac{(p_2 q_3)}{\op1^2}\right) \nonumber \\
  &\hspace*{-0.3cm}-&\hspace*{-0.3cm}
  2(p_2 q_3)+2r_0\op1-\frac{(p_1 q_3)^2}{(q_3\overline{q}_3)}\Bigg(
  2+\frac{r_0}{\op1} \nonumber \\
  &\hspace*{-0.3cm}+&\hspace*{-0.3cm}
  \frac{(p_1 q_3)}{\op1^2}\left(1+\frac{r_0}{2\op1}\right)\Bigg)\Bigg)
  \delta(\sqrt{s}-\op1-\op2-r_0)\nonumber \\
  &\hspace*{-0.3cm}-&\hspace*{-0.3cm}
  \op1^2\Bigg(\frac{(p_1 p_2)}{(p_1 q_3)(p_2 q_3)}
  +\frac{\oq3}{\op1(p_1 q_3)}+\frac{(p_2 q_3)}{2(p_1 q_3)\oq3^2} \nonumber \\
  &\hspace*{-0.3cm}-&\hspace*{-0.3cm}
  \frac{(p_1 q_3)}{2(p_2 q_3)\oq3^2}-\frac{2}{(p_1 q_3)}\Bigg)
  \delta(\sqrt{s}-\op1-\op2). \label{f3def}
  \eeqa
Integrating the singular terms with $D=4-2\eps+c_R 2 \eps$ and
expanding around $\epsilon=0$ we obtain \eqn{eq:W1_A1_Amp_Res}.

\subsection*{Acknowledgments}

DAF acknowledges support from a PPARC studentship.


\end{document}